\documentclass[aps,prb,twocolumn,groupedaddress]{revtex4-1}

\usepackage{amsmath}
\usepackage{graphicx}
\usepackage{amssymb}

\begin{document}


\title{Full analytical solution of finite-length armchair/zigzag nanoribbons}

\author{A. Garc\'ia-Fuente}

\author{D. Carrascal}

\author{G. Ross}

\author{J. Ferrer}
\affiliation{Departamento de F\'{\i}sica, Universidad de Oviedo, E-33007 Oviedo, Spain}
\affiliation{Nanomaterials and Nanotechnology Research Center (CINN) CSIC-Universidad de Oviedo, El Entrego E-33424, Spain}

\date{\today}

\begin{abstract}
Finite-length armchair graphene nanoribbons can behave as one dimensional topological materials, that may show edge states 
in their zigzag-terminated edges, depending on their width and termination. We show here a full solution of Tight-Binding graphene 
rectangles of any length and width that can be seen as either finite-length armchair or zigzag ribbons.  We find exact analytical 
expressions for both bulk and edge eigen-states and eigen-energies. We write down exact expressions for the Coulomb interactions 
among edge states and introduce a Hubbard-dimer model to analyse the emergence and features of different magnetic states at 
the edges, whose existence depends on the ribbon length.  We find ample room for experimental testing of our predictions in
$N=5$ armchair ribbons. We compare the analytical results with {\it ab initio} simulations to benchmark 
the quality of the dimer model and to set its parameters. A further detailed analysis of the {\it ab initio} Hamiltonian allows 
us to identify those variations of the Tight-Binding parameters that affect the topological properties of the ribbons.
\end{abstract}

\maketitle


\section{Introduction} \label{Sec:intro}

The experimental identification of graphene sheets almost two decades ago \cite{Novoselov2004} lead to the development of a 
whole new branch of condensed matter physics, that of 2D materials. Since then, several new 2D materials, such as 
silicene,\cite{Vogt2012} phosphorene \cite{Liu2014} or MoS$_2$ \cite{Wang2012} have been fabricated, presenting different and 
exotic properties. However, the interest in graphene-based structures has not diminished during the years. In particular, graphene 
nanoribbons (GNRs) keep attracting attention due to their characteristic electronic and magnetic properties, usually related to 
the presence of topologically protected edge states around their zigzag terminations. 
Experimentally, bottom-up techniques have enabled the fabrication of long armchair GNRs of different widths and finite length 
from molecular precursors  with atomic precision.\cite{Cai2010,Kimouche2015,Wang2016,Talirz2017,Yamaguchi2020,Way2022} 
The existence of edge states at the zigzag ends of some of these ribbons has been confirmed by scanning tunneling 
microscopy,\cite{Wang2016} while transport measurements have demonstrated their magnetic character.\cite{Lawrence2020} 

From the theoretical point of view, the existence of edge states localized at the zigzag edges of 
GNRs \cite{Nakada1996,Brey2006,Son2006,Yang2007,Jung2009,Fernandez-Rossier2008,Ijas2013} and graphene islands of 
different shapes \cite{Wimmer2010} was predicted long time ago. But, only after the work of Cao {\it et al} in 2017,\cite{Cao2017} the 
topological nature of these edge states has been unveiled. Cao {\it et al} made use of a $Z_2$ topological
invariant that depended on the ribbon width and termination and could be computed by determining the Zak phase from the 
Tight-Binding (TB) wave-functions.\cite{zak1989,Fu2007}  Finite-length armchair ribbons could be classified into a $Z_2=1$ topological 
class, where ribbons host robust edge states, and a $Z_2=0$, topologically trivial class. Furthermore, GNR-based heterostructures were 
proposed and found, where protected edge states emerge at the boundaries between GNRs of different 
topology.\cite{Cao2017, Rhim2017} This work led to a renovated interest in finite-length GNRs and the topological states at their ends, 
with new efforts dedicated to further characterize them both computationally \cite{Lopez-Sancho2021} and 
experimentally.\cite{Rizzo2018,Groning2018}

We analyse here the emergence and features of edge states in finite-length GNRs, where we map the ribbons to a waveguide of 
Schrieffer-Heeger-Su (SSH) \cite{Su1979} transverse modes.  The ribbons that we discuss here can be viewed as either armchair or
zigzag depending on the width/length aspect ratio, or more generally as graphene rectangles or {\it rectangulenes}. We present a 
full analytical solution of a graphene TB Hamiltonian with open boundary conditions in all directions to take into 
account the ribbons finite width and length. We uncover the bulk-boundary condition \cite{laszlobook} by relating the ribbon 
Hamiltonian winding number to the quantization condition for the bulk and edge states.  Our analysis goes beyond a topological 
classification since we are able to characterize fully the edge wave-function spatial distribution, that determines the strength of 
electron-electron interactions and hence the magnetic properties of the ribbons. We also show how and why topological predictions 
for edge states fail for short enough ribbons. 

\begin{figure*}[ht!] 
\centering
\includegraphics[width=\textwidth]{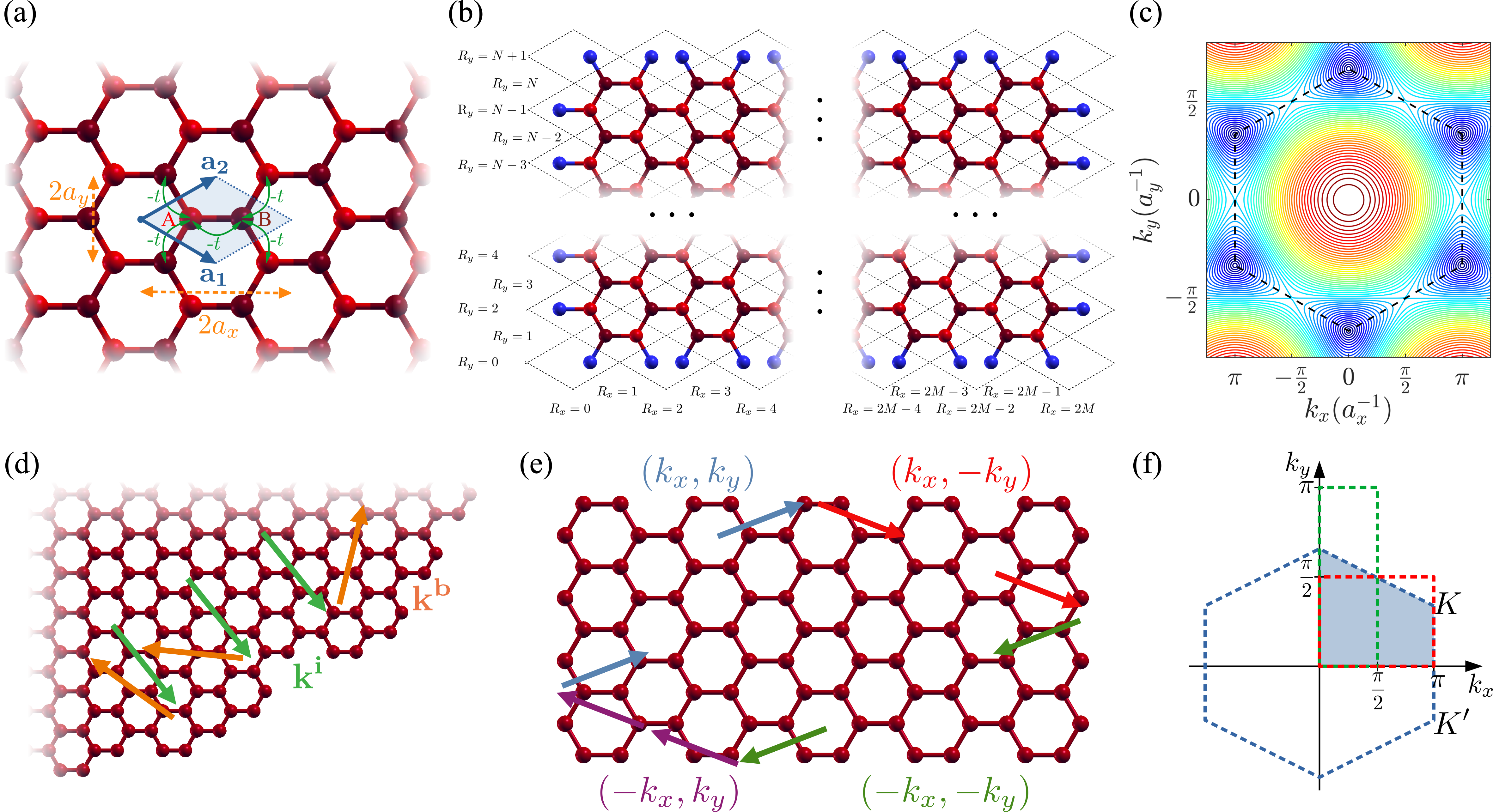}
\caption{
(a) Representation of an infinite graphene sheet, where the $A-$ and $B-$sublattice carbon atoms are shown as light and 
dark red spheres, respectively. The unit cell is defined by the lattice vectors ${\bf a_1}$ and ${\bf a_2}$. 
Hopping integrals between first-neighbour $p_z$ orbitals are shown in green. 
$a_x$ and $a_y$ are lengths used to scale all distances along the X and Y axes, respectively.
(b) Scheme of a finite graphene ribbon characterized by a width $N$ and a length $M$.  Blue spheres represent $fake$ atoms that 
define the boundary conditions. A dotted grid and values of $R_x$ and $R_y$ are depicted to identify the lattice cells and 
XY-components of their corresponding lattice vectors. 
(c) Iso-energy curves of an infinite graphene sheet in the reciprocal space. 
Different colors are used to identify different energies. 
(d) An incident wave with wave-vector ${\bf k^i}$ (green) impinging in a rough edge 
bounces back in several directions indicated by the orange wave-vectors ${\bf k^b}$, which leads to a chaotic cavity.
(e) Incident waves that bounce back in the 
straight edges of a finite ribbon only change the sign of one of their wave-vector components at a time.
(f) Representation of the reciprocal space of graphene. The boundaries of the First Brillouin zone are drawn with blue dashed lines. 
The area marked in shaded blue represents the region of the First Brillouin zone where all the states of the finite ribbon can be folded. 
Green and red dashed lines show other equivalent regions of the 
reciprocal space. We use the red dashed region in this article because $\Delta_y \ge 0$ inside it.
}
\label{fig:multi1}
\end{figure*}

The analytical solution of infinite-length ribbons with armchair and zigzag or arbitrary orientation
has been known for a long time now,\cite{Wakabayashi2010, Delplace2011}
where a band of edge states associated to zigzag-like terminations appears at the lateral edges of the ribbons. Akhmerov and 
coworkers \cite{Akhmerov2011} analysed the nature of edge states in finite-size graphene dots. Little effort has been done however 
in obtaining the analytical solution of finite-length ribbons, where a small set of edge states appears at the ribbon ends rather than 
along the ribbon. In addition,  previous solutions usually required the definition of a one dimensional unit cell having several 
(more than two) basis states to generate the ribbon, while our analysis shows that two orbitals suffice just as in bulk graphene if 
one chooses the adequate boundary conditions. Hence the connection to bulk graphene and to the SSH model is made transparent.

We also deduce a double-site Hubbard model that accounts for the magnetic states of small-width armchair GNR (AGNR), and show how
different magnetic states emerge as the ribbon length increases. We find that the length windows between transitions is large
enough for $N=5$ AGNR to leave ample room for experimental testing.

We complement our analytical TB approach with Density Functional Theory (DFT) simulations to deliver a complete 
theoretical characterization of the ribbons, with the possibility of getting in closer contact to current-day experiments. 
We are therefore able to characterize completely the TB parameters, where we discuss how needed second- and third-neighbour
hopping elements affect the topology of a given ribbon.

The outline of this article is as follows. Section  \ref{Sec:tb} introduces the ribbon TB Hamiltonian, our handling of open boundary conditions
and explains the full analytical solution, together with a complete analysis of the exact bulk and edge states. Section \ref{Sec:dm} 
introduces an effective Hubbard dimer model that accounts for the electron-electron interactions between edge states, whose
parameters are fully determined thanks to the knowledge of the exact wave-functions. The section includes a detailed analysis of the
magnetic mean-field solutions of the model, where their existence is found to depend on the ribbon length.
Section \ref{Sec:dft} compares the analytical results to {\it ab initio} DFT simulations of the ribbons. A close inspection and handling of
the DFT Hamiltonian allows us to map it to a third-nearest neighbour TB Hamiltonian. We discuss how the extra neighbour terms affect 
the robustness of the edge states.  Section \ref{Sec:conclusions} summarizes our
main conclusions. Appendix A delivers a pedagogical description of the analytical solution of finite-length one dimensional chains.  Appendix B
shows our DFT results for $N=7$ and $9$ AGNR.


\section{Analytical solution of finite-length GNRs} \label{Sec:tb}

\subsection{ Hamiltonian, eigen-states and eigen-functions of an infinite graphene sheet}
We discuss here shortly the solution of an infinite graphene sheet to introduce notation that will help us to discuss the finite-length case.
We consider the primitive unit cell depicted in Fig. \ref{fig:multi1} (a), where we consider a single $p_z$ orbital per carbon atom as usual. 
Lattice vectors ${\bf R}$ are spanned in terms of the primitive vectors ${\bf a}_1$ and ${\bf a}_2$. Distances along the X and Y axes
are measured in units of the primitive vector components lengths  $a_x$ $\simeq$ 2.13 \AA\ and $a_y$ $\simeq$ 1.23 \AA.
This choice simplifies the algebraic expressions below,  rendering our results independent of uniform distortions of the lattice from the hexagonal structure (however notice that lattice distortions affect the value of the hopping integrals, as we will discuss later). 
Then, the Hamiltonian of the system can be written as follows:
\begin{equation}
  \label{eq:hamiltonian}
  \begin{aligned}
     \hat{\cal{H}}= & -t\sum_{\bf{R}}\sum_{\bf{ \delta}={\bf 0},{\bf a_1},{\bf a_2}} 
     \left ( \hat{a}_{\bf R}^{\dagger}\,\hat{b}_{{\bf R}-{\bf \delta}}^{} +  \hat{b}_{\bf R}^{\dagger}\,\hat{a}_{{\bf R}+{\bf \delta}} ^{} \right)
   \end{aligned}
\end{equation}

\noindent
where $\hat{a}_{\bf R}^{\dagger}$ ($\hat{b}_{\bf R}^{\dagger}$) and $\hat{a}_{\bf R}^{}$ ($\hat{b}_{\bf R}^{}$) are the creation 
and annihilation operators acting on site 
$A$ ($B$) of the unit cell defined by the lattice vector $\bf{R}$.  We are considering only nearest-neighbors hopping integrals $-t$ 
(Fig. \ref{fig:multi1} (a)) and set all on-site energies to zero. We gather the basis states centered at sites 
$A$ or $B$ of each $\bf{R}$ unit cell into a vector
\begin{equation}
{\bf |R\, \rangle}=\left(\begin{matrix} |{\bf R},A \rangle\\ |{\bf R},B \rangle\end{matrix} \right )
\end{equation}
Then, any eigen-state wave-function can be written as the linear combination
\begin{equation}
\label{eq:GeneralWF}
 |\Psi \rangle = \sum_{\bf R}  C_{\bf R}^{\top}\,{\bf |R\, \rangle}
\end{equation}
where translational symmetry dictates that the Bloch coefficients must be decomposed as  
\begin{equation}
C_{\bf R}=\left(\begin{matrix} \,c_{\bf R}^A \\ \,c_{\bf R}^B\end{matrix}\right)=e^{i\bf{k}\,\bf{R}}\,C_{\bf k} = e^{i\bf{k}\,\bf{R}}\,\left(\begin{matrix} \,c_{\bf k}^A \\ \,c_{\bf k}^B\end{matrix}\right)
\label{eq:blochC}
\end{equation}
The wave-vectors {\bf k} label the Bloch eigen-states. They must be real to guarantee that the 
wave-function is normalizable, and are determined by imposing suitable (periodic) boundary conditions.  
The 2$\times$2 Hamiltonian can be written as:
\begin{equation}
\label{eq:bulkH}
H = \left ( \begin{matrix} 0 & -f^{*}(\bf{k})\\ -f(\bf{k}) & 0 \end{matrix} \right ) 
= - \left | f(\bf{k}) \right | \left ( \begin{matrix} 0 & e^{-i\theta_{\bf{k}}} \\ e^{i\theta_{\bf{k}}} & 0 \end{matrix} \right )
\end{equation}
where
\begin{eqnarray}
f({\bf k})&=& t\left ( 1 + e^{i\bf{k}\bf{a}_{1}} + e^{i\bf{k}\bf{a}_{2}} \right )= t\left ( 1 + \Delta_y e^{i k_x} \right ) \\
|f({\bf k})| &=& t \sqrt{1+\Delta_y^2 + 2\,\Delta_y\,\cos{\left ( k_x \right ) }}
\end{eqnarray}

\noindent
and $\theta_{\bf k}$ is the polar angle of $f({\bf k})$.
We have dumped all the $k_y$ dependence  into the function $\Delta_y=\Delta(k_y)=2\,\cos{\left ( k_y \right )}$ that depends 
only on the modulus of $k_y$. 
It is now straightforward to see that the eigen-values  and  eigen-function coefficients can be written as follows:
\begin{equation}
\label{eq:sheetsol}
\begin{aligned} 
&\varepsilon_{{\bf k}\,\tau} = -\tau \left | f(\bf{k}) \right |  \\
&C_{{\bf k} \, \tau}= \frac{1}{\sqrt{2}}\,\left(\begin{matrix} 1\\ \tau e^{i\theta _k}\end{matrix}\right)
\end{aligned}
\end{equation}
where $\tau=\pm$ labels graphene's valence and conduction bands. 

\subsection{Open boundary conditions in a finite armchair ribbon}

We consider now armchair nanoribbons of finite length, defined by their width $N$ (e.g.: the number of atomic rows) and their length $M$ 
(e.g.: the number of hexagons along the length of the ribbon) as shown in Fig. \ref{fig:multi1} (b). We focus on odd values of $N$ because 
those are the kind of ribbons that can be obtained experimentally. However, most of our analytical results are also valid for an even value of $N$, 
and we also comment briefly those cases in the following sections.
In contrast with the infinite sheet, translational symmetry is broken now because edge atoms exist that have a coordination number of two 
instead of three.  We can however restore translational symmetry by inserting fake atoms at the edges as drawn in 
Fig. \ref{fig:multi1} (b), so that edge atoms recover a coordination number of three.  The cost for doing so consists of inserting extra 
equations that ensure
that the wave-function is exactly zero at the fake-atom positions. These extra equations are the finite-length boundary conditions
that replace the periodic boundary conditions of the infinite sheet.

We note now that the Bloch coefficients $C_{\bf R}$ in Eq. (\ref{eq:blochC}) are non-zero for all ${\bf R}$, so that they cannot meet the 
boundary condition equations. 
We can however take advantage of the fact that any linear combination of same-energy bulk coefficients $C_{\bf k}$ is also an 
eigen-state of the system with the same energy. We therefore search for those linear combinations that fulfill the boundary conditions.  
Graphene bulk eigen-states have large degeneracies at most energies. This is illustrated in  Fig. \ref{fig:multi1} (c), 
where isoenergy curves within graphene's bulk Brillouin zone are drawn. The set of possible linear combinations can 
however be restricted by noticing that it is the edges that mix waves as we illustrate in Figs.  \ref{fig:multi1} (d) and (e).  
Indeed, any wave with wave-vector ${\bf k^i}$ that impinges on an edge must bounce back with a momentum 
${\bf k^b}$ whose components satisfy  $k^b_\parallel=k^i_\parallel$ and  $k^b_\perp=-k^i_\perp$. 
Fig. \ref{fig:multi1} (d) shows a wave inpinging on an edge that has irregular shape, typical of a chaotic cavity. This edge 
gives rise to many outgoing waves, and all of them must be included in the linear combination. In contrast, Fig. \ref{fig:multi1} (e) shows
equal-energy waves inside one of the ribbons that we study in this article.  Then the edges' symmetries restrict the 
possible linear combinations to just four waves for each incident wave-vector ${\bf k^i}$. We denote the set of 
four waves by ${\bf k}_{\sigma,\sigma'}= (\sigma k_x,\sigma' k_y)$, where $\sigma,\,\sigma'=\pm$.
These considerations imply that the wave-function coefficients consist of the summation of four Bloch coefficients:
\begin{equation}
C_{\bf R} = \sum_{\sigma,\sigma'=\pm}\,A_{\sigma,\sigma'} e^{i {\bf k}_{\sigma,\sigma'} {\bf R}}\,C_{{\bf k}_{\sigma,\sigma'}}
\end{equation}
with boundary conditions 
\begin{equation}
\label{eq:obc1}
\begin{aligned} 
c_{(R_x,R_y=0)}^{A} = c_{(R_x,R_y=N+1)}^{A} =  0; \ \ & R_x=2,4...2M \\
c_{(R_x,R_y=0)}^{B} = c_{(R_x,R_y=N+1)}^{B} =  0; \ \ & R_x=0,2,...2M-2 \\
c_{(R_x=0,R_y)}^{A} = c_{(R_x=2M,R_y)}^{B} =  0; \ \ & R_y=2,4...N-1
\end{aligned}
\end{equation}

\begin{figure*}[ht!] 
\centering
\includegraphics[width=\textwidth]{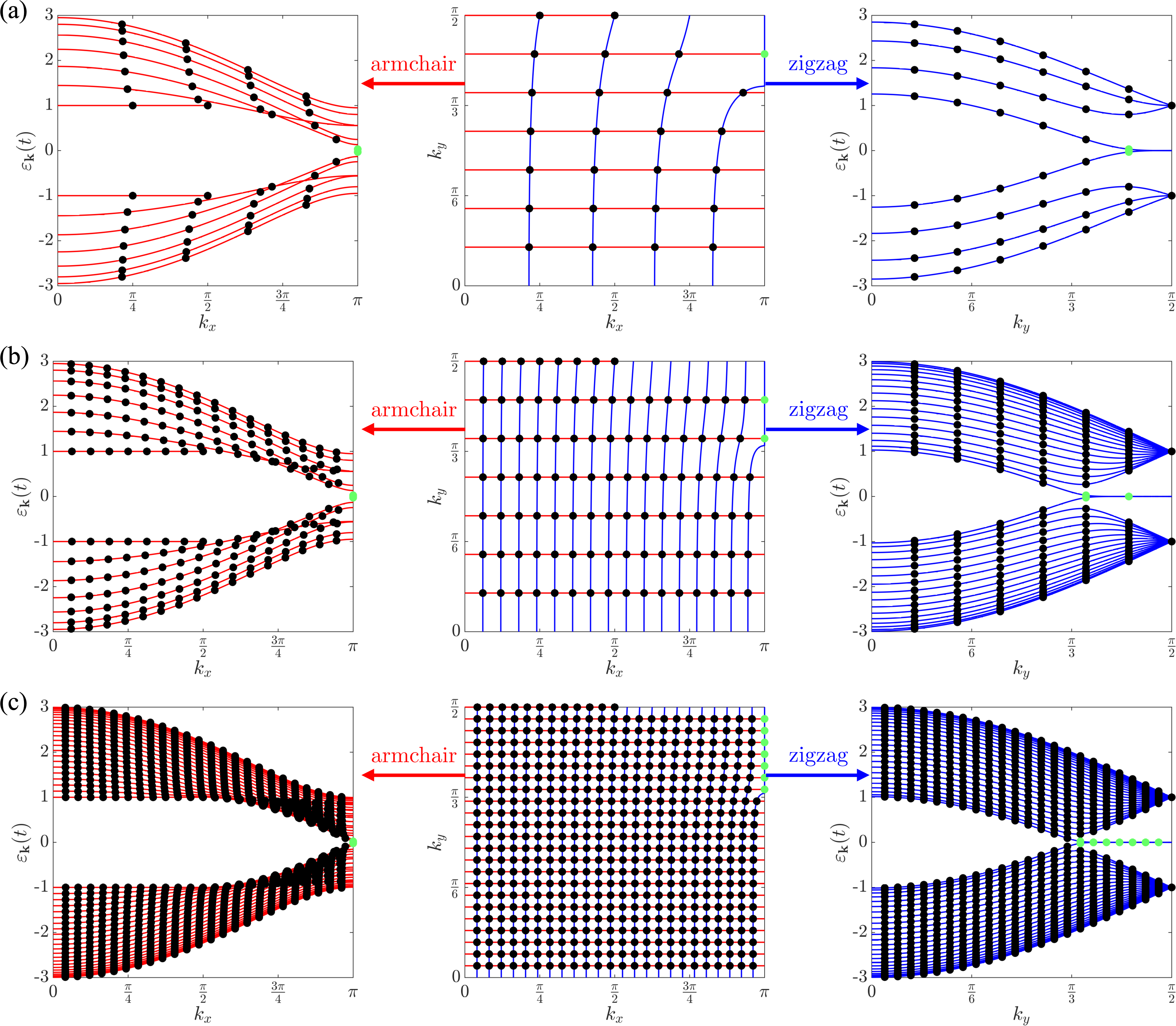}
\caption{Grid of quantized solutions $\left ( k_{x,\,\alpha\,\beta}, k_{y,\,\alpha} \right )$ of a graphene rectangulene with (a) $N=9$, $M=2$; 
(b) $N=9$, $M=8$; (c) $N=45$, $M=12$ ribbon. Each central panel shows the grid where black and green dots represent the real 
$k_{x,\,\alpha\,\beta}$ (bulk states) and complex $\pi-i\,q_\alpha$ (edge states) solutions. Red and blue lines 
are an eye guide to visualized correlations. The quantized eigen-energies can be visualized from the band structure of an infinite 
armchair (left panel) or zigzag (right panel) ribbon.}
\label{fig:ribbands}
\end{figure*}

The eigen-functions of the system are characterized by a single wave-vector ${\bf k}$ that
lies inside a region within the first quadrant of the Brillouin Zone region. We draw in Fig. \ref{fig:multi1} (f) several 
possible choices for the region. We have chosen the region enclosed by the dashed red lines 
($k_x\in \left [ 0, \pi \right ], k_y\in \left [ 0, \frac{\pi}{2} \right ]$) because the function $\Delta_y \ge 0$ inside it. 
To proceed, we notice that the Bloch coefficients $C_{\bf k}$ in Eq.  (\ref{eq:sheetsol}) depend only on the modulus of $k_y$, 
e.g.: $C_{(kx,ky)}=C_{(kx,-ky)}$ and we denote these by $C_{k_x}$ below.  This means that we can factorize the wave-function 
coefficients as follows:

\begin{eqnarray}
\label{eq:crjbulk1}
C_{\bf{R}} &=& D_{R_x}(k_x, k_y)\, E_{R_y}(k_y) \\
\label{eq:crjbulk3}
D_{R_x}(k_x,k_y) &=&  D_+\, e^{ik_x\,R_x}\,C_{k_x} + D_-\, e^{-ik_x\,R_x}\,C_{-k_x} \\
\label{eq:crjbulk2}
E_{R_y}(k_y) &=& E_+ \,e^{ik_y\,R_y} + E_- \,e^{-ik_y\,R_y}  
\end{eqnarray}

Notice that $D_{R_x}(k_x,k_y)$ is a vector of components $\left ( d_{R_x}^A, d_{R_x}^B \right )$, while $E_{R_y}(k_y)$
is just a scalar.
Similarly, the boundary conditions in Eq. (\ref{eq:obc1}) can be written in a factorized form as follows:
\begin{eqnarray}
\label{eq:bc2}
d_{R_x=0}^A &=& d_{R_x=2\,M}^B = 0 \\
\label{eq:bc1}
E_{R_y=0}&=&E_{R_y=N+1}=0
\end{eqnarray}
Eqs. (\ref{eq:crjbulk1}) through (\ref{eq:bc1}) are the first central result of this article. We can infer from them that a finite-length armchair 
GNR system is a separable problem in the sense that it can be decomposed into two much simpler finite-length one-dimensional 
models as follows.  

Eqs. (\ref{eq:crjbulk2}) and (\ref{eq:bc1}) correspond to a simple $N$-site mono-atomic chain that lies 
along the Y-direction. As also shown in Appendix \ref{Appendix:obc}, the boundary condition of Eq. (\ref{eq:bc1}) quantizes 
the $k_y$ wave-vectors as follows:
\begin{equation}
\label{eq:kycondition}
\sin{ \left ( (N+1) k_y \right )}=0 \Rightarrow k_{y}=k_\alpha=\pi\,\frac{\alpha}{N+1}
\end{equation}
where we have labeled the allowed  wave-vectors by the integer number $\alpha$, with $\alpha = 1,\, ...,\, \frac{N+1}{2}$. These $k_\alpha$ 
wave-vectors lie all inside the ribbon Brillouin zone shown in red lines in Fig. \ref{fig:multi1} (f). 

The quantized $k_\alpha$ wave-vectors enter Eqs. (\ref{eq:crjbulk3}) and (\ref{eq:bc2}) as a parameter through the function 
$\Delta(k_y)$, and we call $\Delta_\alpha=\Delta(k_\alpha)=2\cos{\left ( k_\alpha \right )}$ to simplify the notation below. Then, these two equations correspond 
to a set of dimerized chains lying along the X-axis that have $2\,M$ cells. Each of the chains correspond to a 
different $\Delta_\alpha$. The dimerized TB chain is solved in detail in Appendix \ref{Appendix:obc}.
The boundary conditions of Eq. (\ref{eq:bc2}) fix the $k_{x,\,\alpha}$  allowed values for each $k_{y,\,\alpha}$ via the equation

\begin{equation}
\sin{ \left ( 2M k_{x,\,\alpha} +\theta_{\bf k} \right ) }=0 \Rightarrow 2M k_{x,\,\alpha} + \theta_{\bf k} = \beta \pi
\label{eq:kxcondition}
\end{equation}

\noindent
with the additional condition that the wave-vectors must lie within the ribbon Brillouin Zone, $k_{x,\,\alpha}\in \left [ 0, \pi \right ]$.
For each given value of $\alpha$, the integer value $\beta$ univocally defines the value of $k_x$, so we label
$k_{x,\,\alpha,\,\beta}=k_{\alpha\,\beta}$.
We define a critical $\Delta_{\alpha}^c=1$, that corresponds to a critical wave-vector $k_{\alpha}^c=\pi/3$.
Then, the above equation has $2\,M$ real solutions so that $\beta=1,\, ...,\,2\,M$ if (a) $\Delta_\alpha > \Delta_{\alpha}^c $ ($k_\alpha<\pi/3$), 
or (b) if $\Delta_\alpha < \Delta_{\alpha}^c$ ($k_\alpha>\pi/3$) and the chain length $M<M_c=\frac{\Delta_\alpha}{2(1-\Delta_\alpha)}$. 
However, the above equation has only $2\,M-1$ real solutions if $\Delta_\alpha < \Delta_{\alpha}^c$ ($k_\alpha>\pi/3$) and the chain 
$M>M_c$, so that $\beta=1,\, ...,\,2\,M-1$ in this case.  
The missing solution can be found by setting $k_{x,\,\alpha}=\pi-i\,q_\alpha$ where $q_\alpha$ is determined by the equation

\begin{equation}
\sinh{ \left ( 2 M q_\alpha + \overline{\theta}_q \right ) } = 0 \rightarrow 2 M q_\alpha + \overline{\theta}_q  = 0
\end{equation}

\noindent
where we have introduced $\overline{\theta}_q$ in analogy to $\theta_{\bf k}$ as:

\begin{equation}
e^{2\overline{\theta}_q}=\frac{f\left( \pi - iq,k_\alpha \right ) }{f \left ( \pi+iq, k_\alpha \right )}
\end{equation}

The case of $k_\alpha = \pi$ is especial. In that case $\Delta_\alpha=0$, but $E_{R_y}=0$ for all even values of $R_y$. Therefore, there is no
condition over $k_x$, instead we obtain $M$ degenerated states of $\varepsilon_{\bf k}=-\tau\,t$. However, we can still use condition
(\ref{eq:kxcondition}) to obtain these $M$ bulk states in the range $k_x \in [0, \frac{\pi}{2}]$.

The central panels in Fig. \ref{fig:ribbands} show the resulting grid of real $(k_{x,\,\alpha,\,\beta},\,k_{y,\,\alpha})=(k_{\alpha\,\beta},\,k_{\alpha})$ 
solutions within the ribbon Brillouin Zone for ribbons of two selected widths.  
In these central panels, the quantization condition (\ref{eq:kycondition}) is represented by red horizontal lines, while blue lines represent 
the quantization condition (\ref{eq:kxcondition}). The curvature of the latter represents the dependence of quantized $k_x$ values in $k_y$.
The last blue curve hits $k_x = \pi$ at $k_y > k_{\alpha}^c$, where complex values of $k_x$ arise. Considering only one of the quantization
conditions we recover the band structure of armchair (left) or zigzag (right) ribbons. Considering both conditions we obtain a grid of points that 
represent the actual $(k_{\alpha \beta}, k_{\alpha})$ states of the ribbon. Fig. \ref{fig:ribbands} (a) and (b) show how the number of edge states
(represented by green dots) of a ribbon of a given width depends on its length. These edge states are part of the zigzag band structure, but
fall inside the energy gap of the armchair band structure. 
The number of edge states of the ribbon is given by the number of allowed $k_\alpha \in (\pi/3,\,\pi/2)$,  which gives
$\mathrm{floor}(\frac{N+1}{6})$ for odd values of $N$. Each putative edge state must also fulfill the extra condition $M>M_c$.

The bulk eigen-states  have wave-functions and eigen-energies given by

\begin{widetext}
\begin{equation}
|\Psi_{\alpha\,\beta\,\tau}\rangle=\sqrt{\frac{8}{(N+1)\,A_{\alpha\,\beta}}}\,
\sum_{\bf{R}}\,\sin{ \left ( k_\alpha\,R_y \right ) }\, 
\left(\begin{matrix}
\sin{\left ( k_{\alpha\,\beta}\,R_x \right ) }\\
\tau\,(-1)^{\beta+1}\sin{ \left ( k_{\alpha\,\beta}\,\left ( 2M-R_x \right ) \right ) }  
\end{matrix}\right)^{\top} |\,{\bf R}\rangle
\end{equation}
\end{widetext}

\begin{eqnarray}
A_{\alpha\,\beta}&=& {\cal M}-\frac{\sin{\left ( {\cal M}\,k_{\alpha\,\beta} \right ) }}{\sin{ \left ( k_{\alpha\,\beta} \right ) }}\\
\frac{\varepsilon_{\alpha\,\beta\,\tau}}{t}&=&-\tau\,\sqrt{1+\Delta_\alpha^2+2\,\Delta_\alpha\,\cos{\left ( k_{\alpha\,\beta} \right )}}
\end{eqnarray}

\noindent
where $\tau=\pm$,  and we have used the shorthand ${\cal M}=4\,M+1$, while the edge eigen-states wave-functions and eigen-energies are

\begin{widetext}
\begin{equation}
|\Psi_{\alpha\,\tau}\rangle=\sqrt{\frac{8}{(N+1)\,B_{\alpha}}}\,
\sum_{\bf R}\,\sin{ \left ( k_\alpha\,R_y \right ) }\,\left(\begin{matrix} (-1)^{R_x} \sinh{ \left ( q_\alpha\,R_x \right ) }
\\\ \tau\,(-1)^{R_x+1}\sinh{ \left ( q_\alpha\,\left ( 2M-R_x \right ) \right ) }\end{matrix}\right)^\top |\,{\bf R}\rangle\\
\end{equation}
\end{widetext}

\begin{eqnarray}
B_\alpha&=&\frac{\sinh{\left ( {\cal M}\,q_\alpha \right ) }}{\sinh{ \left ( q_\alpha \right )}}-{\cal M}\\
\label{eq:tbfinaleqs}
\frac{\varepsilon_{\alpha\,\tau}}{t}&=&-\tau \sqrt{1+\Delta_\alpha^2-2\Delta_\alpha\, \cosh{\left ( q_\alpha \right )}}\\
& =& -\tau  \frac{\sinh{\left ( q_\alpha \right ) }}{\sinh{\left ( \left ( 2M+1 \right )\, q_\alpha \right )}}\nonumber
\end{eqnarray}

All these results are valid for both odd and even values of $N$. The only noticeable difference is that the especial case of $k_\alpha = \pi$ only
appears for ribbons with odd $N$, and that in this case the number of edge states of the ribbon is given by $\mathrm{floor}(\frac{N+4}{6})$.
Eqs. (\ref{eq:kycondition}) through (\ref{eq:tbfinaleqs}) give the full solution of the TB finite length nanoribbon and are the second central result of this article.

\subsection{Number of edge states and topology}

We note the well-known fact that SSH chains can be classified according to two topological categories depending on the ratio between their hopping integrals.
In the correspondence between the GNR along the X direction and the dimerized chain, this ratio is just $\Delta_\alpha^c$. SSH chains with 
$\Delta_\alpha>\Delta_\alpha^c$ are topologically trivial in the sense that they host only bulk states. SSH chains with $\Delta_\alpha<\Delta_\alpha^c$
are topological, they host topologically protected edge states (beyond a certain length). Appendix \ref{Appendix:obc} 
shows in detail the content of the bulk/boundary principle for SSH chains. 

\begin{figure}[ht!] 
\centering
\includegraphics[width=1.00\columnwidth]{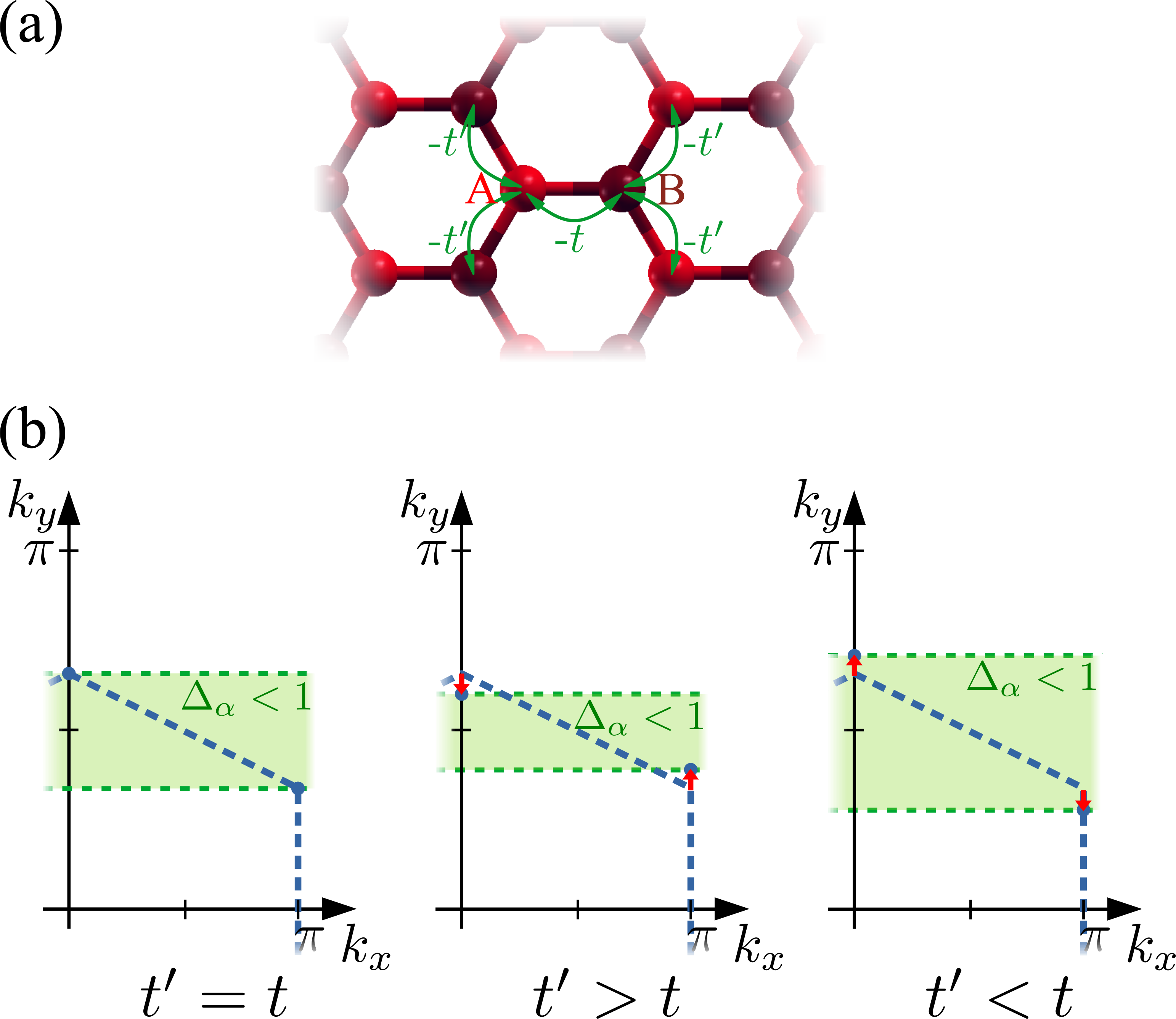}
\caption{(a) Sketch of the modified TB approach with 2 different hopping integrals $t$ and $t'$. (b) Having $t'\neq t$ impacts on the
size of the allowed region of the ribbon BZ where edge states appear, because changing $t'$ shifts the Dirac point up and down.}
\label{fig:t-tp}
\end{figure}

We can separate armchair ribbons into 3 groups, corresponding to $N=3p$, $3p+1$ or $3p+2$. For
long enough ribbons with an odd value of $N$, $N=3p$ contains $\frac{p-1}{2}$ edge states
for each edge, while $N=3p+1$ contains $\frac{p}{2}$. Within this approach $N=3p+2$ infinite ribbons are found 
to be metallic, as the $k_\alpha=\frac{\pi}{3}$ band passes through the Dirac point $K$.
However, DFT results show that these ribbons have a small gap.\cite{Yang2007} A modification of our TB model that has
different hopping integrals $-t$ and $-t'$ in the longitudinal and transverse directions of the ribbon  (see Fig. \ref{fig:t-tp})
reproduces this behavior. We redefine $\Delta_\alpha=2 \frac{t'}{t} \cos{\left ( k_\alpha \right )}$, so that the rest 
of the problem remains the same.

Then, if $t'>t$, the region of reciprocal space that represents $\Delta_\alpha < 1$ is reduced and the $N=3p+2$ ribbons can 
only present $\frac{p-1}{2}$ edge states.
If $t'<t$, the same region is increased
and these ribbons can present $\frac{p}{2}$ edge states. This is shown in Fig. \ref{fig:t-tp}.

Cao {\it et al} reported the values of the $Z_2$ topological invariant of these infinite ribbons with closed edges,\cite{Cao2017} indicating
$Z_2=\frac{1+\left(-1 \right )^{\left \lfloor \frac{N}{3} \right \rfloor +\left \lfloor \frac{N+1}{2} \right \rfloor}}{2}$. 
$Z_2=1$ (0) is equivalent to a topologically protected odd (even) number of edge states. This is consistent with our results
if $t'<t$. Cao {\it et al} also reported $Z_2$ values for ribbons with open edges, reporting
for them $Z_2=\frac{1-\left(-1 \right )^{\left \lfloor \frac{N}{3} \right \rfloor +\left \lfloor \frac{N+1}{2} \right \rfloor}}{2}$,
that corresponds to the opposite value of $Z_2$ from that of ribbons of the same width $N$ and closed edges, as those analyzed here. 
The analytical
solution of these new ribbons is very similar to that presented here, but in this case condition (\ref{eq:bc2}) must be satisfied for
odd values of $R_y$, and that is not immediately satisfied in $k_\alpha=\pi$ as for the ribbons with closed edges. This leads to an extra couple
of edge states in the limit $\Delta_\alpha=0$, fully localized on the edge atoms and with $\varepsilon_{\alpha \tau} = 0$. 

For ribbons with an even value of $N$, $N=3p$ contains $\frac{p}{2}$ edge states, $N=3p+1$ contains $\frac{p+1}{2}$, while $N=3p+2$ are again metallic
if $t=t'$. Cao {\it et al}\cite{Cao2017} also predicted $Z_2=\frac{1-\left(-1 \right )^{\left \lfloor \frac{N}{3} \right \rfloor +\left \lfloor \frac{N+1}{2} \right \rfloor}}{2}$
for these ribbons. This is again consistent with our calculations if $t'<t$, where we obtain $\frac{p+2}{2}$ edge states for $N=3p+2$ ribbons.

\section{Hubbard Dimer model for inter-edge Coulomb interactions}\label{Sec:dm}

Some of the most relevant features of graphene nanoribbons such as their magnetic, electrical or optical properties originate from the 
strong electron-electron interactions existing among edge states, that go beyond the single-electron picture described above. 
We drop in this section the bulk states and set up a model of interacting edge electrons for the case where we have a single edge-state solution $q_\alpha$.

\subsection{Left and Right edge states}
\begin{figure}[ht!] 
\centering
\includegraphics[width=1.00\columnwidth]{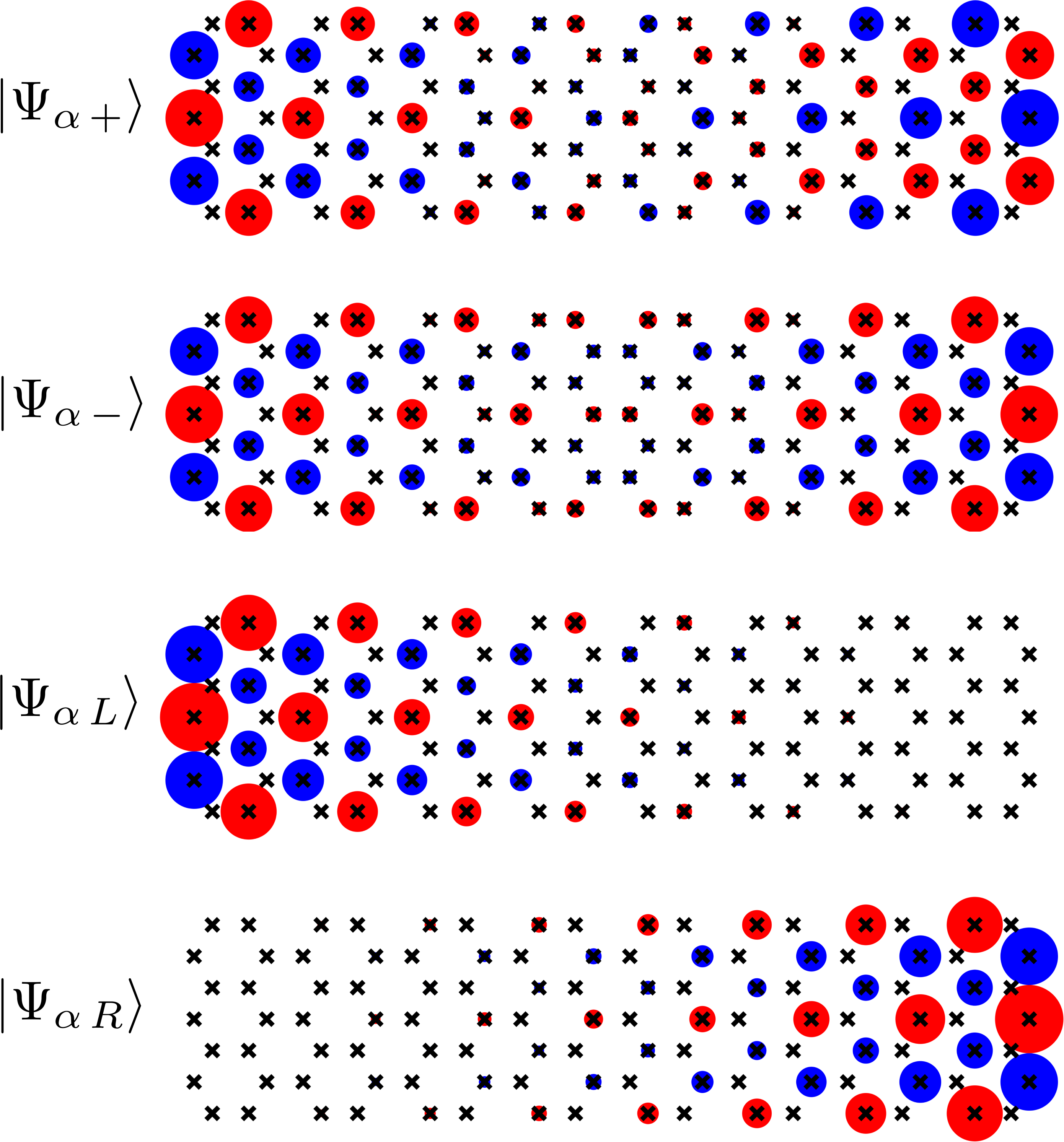}
\caption{Spatial representation of the $\left | \Psi_{\alpha\,+} \right \rangle$, $\left | \Psi_{\alpha\,-} \right \rangle$, 
$\left | \Psi_{\alpha\,L} \right \rangle$ and $\left | \Psi_{\alpha\,R} \right \rangle$ edge states, for a $N=7$, $M=8$ ribbon. Black crosses 
indicate the carbon atom positions. Red (blue) circles at each of those sites indicate positive (negative) values of the 
wave-function coefficients, where the circle radii are proportional to the magnitude of the coefficient.}
\label{fig:phiplot}
\end{figure}

The above $ |\Psi_{\alpha\,\tau} \rangle $ edge eigen-states are delocalized over both edges and both A and B sub-lattices as we show
in Fig. \ref{fig:phiplot}. But we can define alternatively orthogonal zero-energy states that are located at either the left/B 
or right/A edge/sublattice (but are not eigen-states) as follows:

\begin{equation}
\begin{aligned}
\label{eq:edgestates}
|\Psi_{\alpha\, (L,R)} \rangle = \frac{1}{\sqrt{2}}\left ( |\Psi_{\alpha\,-} \rangle \pm |\Psi_{\alpha\,+} \rangle \right )
\end{aligned}
\end{equation}

Alternatively, $ |\Psi_{\alpha\,\tau} \rangle $ can be viewed as the bonding and antibonding states formed by the interaction between 
the single-edge states $|\Psi_{\alpha\,L} \rangle$ and $|\Psi_{\alpha\,R} \rangle$ via an effective hopping integral $t_\alpha$:

\begin{equation}
\label{eq:tp}
\frac{t_\alpha}{t} = \frac{| \varepsilon_{\alpha\,\tau}|}{t} =\frac{\sinh{\left ( q_\alpha \right )} }{\sinh{\left ( \left ( 2M+1 \right )\, q_\alpha \right )}}
\xrightarrow[M\,q_\alpha \gg 1]{} (1 - \Delta_\alpha^2)\, \Delta_\alpha^{2M}
\end{equation}

\begin{figure*}[ht!] 
\centering
\includegraphics[width=1.00\textwidth]{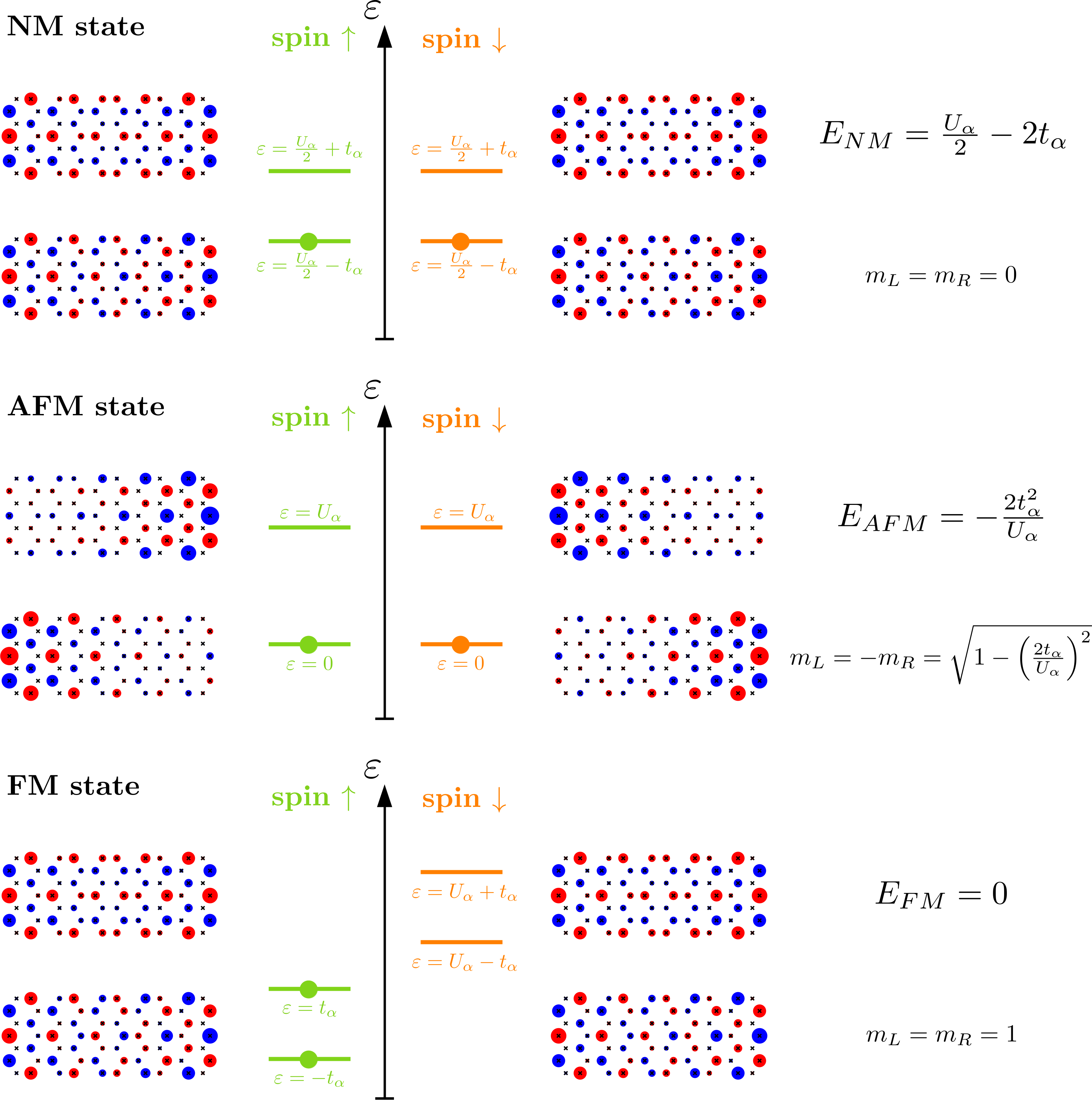}
\caption{Graphical summary of the different mean field magnetic solutions of the Hubbard dimer model at half filling of the states. 
Orange and green lines in the central panel show the one-electron eigen-energies for each solution, where occupied states are 
called HOMO states and are indicated by a dot, while empty states are called LUMO and do not have a dot. The left and right panels 
show the wave-function shapes for a $N=7$, $M=5$ ribbon. Mean field total energies 
and magnetic moments in units of $\mu_B$ are written at the far right side of the figure.}
\label{fig:hubbard}
\end{figure*}

\subsection{Hubbard Dimer model for inter-edge Coulomb interactions}
\begin{figure}[tb!] 
\centering
\includegraphics[width=1.00\columnwidth]{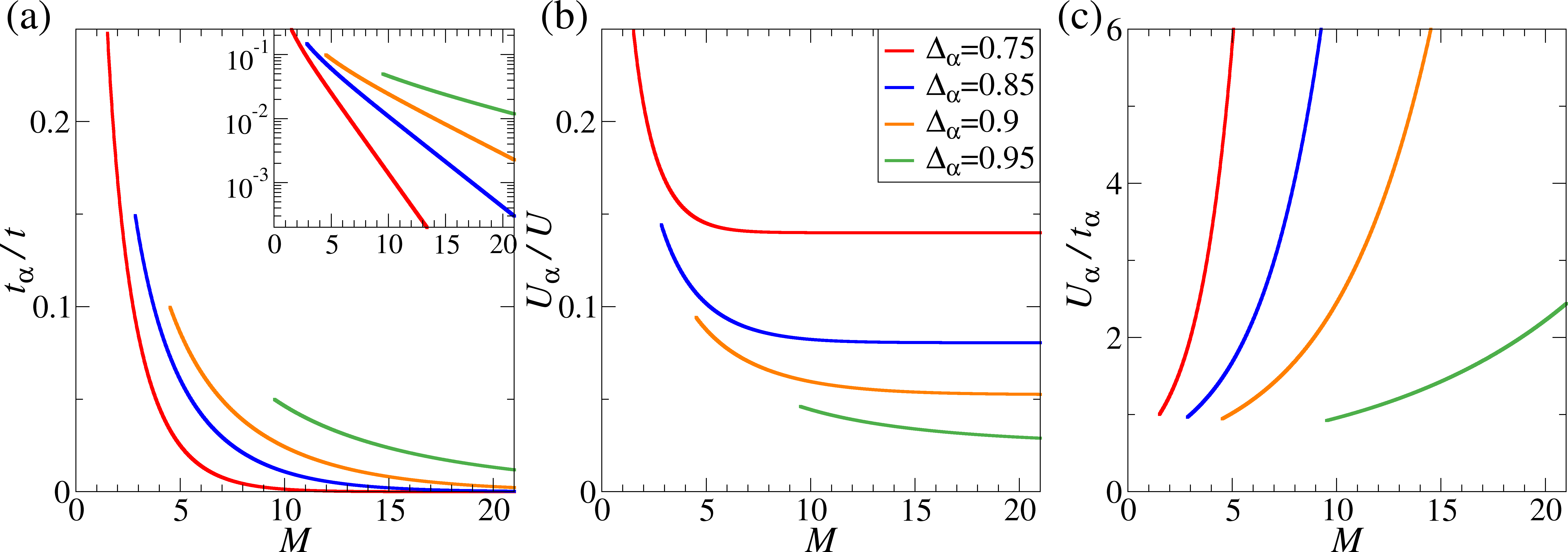}
\caption{(a) $t_\alpha/t$, (b) $U_\alpha/U$ and (c) $U_\alpha / t_\alpha$ (in units of $t/U$) as a function of the length $M$ for an AGNR with $N=5$. The different curves show the results for 
different values of of $\Delta_\alpha$. The inset in figure (a) shows $t_\alpha/t$ in logarithmic scale to highlight its exponential decay.}
\label{fig:Utedge}
\end{figure}

We assume now that electrons in a graphene ribbon obey the Hubbard model to a good approximation. 
One can then show that two electrons in the same single-edge state $q_\alpha$ have opposite spins. We find that their dynamics can be
described to a good approximation by the following Hubbard dimer model

\begin{equation}
\hat{\cal{H}}_\alpha = t_\alpha \sum_\sigma \left ( \hat{c}_{L\sigma}^{\dagger}\hat{c}_{R\sigma} + 
\hat{c}_{R\sigma}^{\dagger}\hat{c}_{L\sigma} \right) + U_\alpha \left 
( \hat{n}_{L\uparrow} \hat{n}_{L\downarrow} + \hat{n}_{R\uparrow} \hat{n}_{R\downarrow} \right )
\label{eq:effectiveH}
\end{equation}

\noindent
where $\hat{c}_{i \sigma}^{\dagger}$ and $\hat{c}_{i \sigma}$ are the spin-$\sigma$ creation and annihilation operators acting on 
the edge states at the $i=L,\,R$ edge,  and $\hat{n}_{i \sigma}$  are their corresponding number operators. The Hubbard-$U$ parameter
is given by 

\begin{eqnarray}
U_\alpha&=& \langle \Psi_{\alpha i\uparrow}|\,\otimes\,\langle\Psi_{\alpha i\downarrow}\,| \,\hat{U}\,
|\Psi_{\alpha i\downarrow}\rangle\,\otimes\,|\Psi_{\alpha i\uparrow}\rangle\\
\label{eq:Up}
&=& \frac{3 \left ( \frac{\sinh{\left ( 2{\cal M}q_\alpha \right )}}{\sinh{ \left ( 2q_\alpha \right )}} -4\frac{\sinh{\left ( {\cal M} q_\alpha \right ) }}{\sinh{\left ( q_\alpha \right ) }} +3 {\cal M} \right )}{(N+1) \left ( \frac{\sinh{ \left ( {\cal M}q_\alpha \right ) }}{\sinh{\left ( q_\alpha \right )}} - {\cal M} \right )^2 } \\
&&\xrightarrow[M\,q_\alpha \gg 1]{}  \frac{3\left ( 1 - \Delta_\alpha^2\right )}{\left ( N+1 \right )  \left ( 1 + \Delta_\alpha^2\right )}U = U_\alpha^0
\end{eqnarray}

\noindent
where $U$ is the local interaction within one atom.
We show in Fig.  \ref{fig:Utedge} (a) and (b) the dependence of $t_\alpha / t$ and $U_\alpha / U$ with the ribbon length $M$ for a $N=5$ ribbon and
different possible values of $\Delta_\alpha$. We find that $t_\alpha/t$ decays exponentially to zero with $M$. In contrast, the Hubbard $U_\alpha/U$ parameter 
decreases strongly for short ribbons, but then levels off and converges to a constant value $U_\alpha^0$ as each edge state adquires its maximum delocalization. 

\subsection{Mean field analysis at half-filling}
We perform a mean field treatment of the Hamiltonian, where we denote 
$n_{i \sigma}=\langle\hat{n}_{i \sigma}\rangle$.  We also denote by $m_i = n_{i \uparrow} - n_{i \downarrow}$ 
the magnetic moment in units of $\mu_B$ at either the $i=L$ or the $i=R$ ribbon edge. We shall restrict the analysis
to the half-filled case so that $n_{L\uparrow}+n_{R\uparrow}+n_{L\downarrow}+n_{R\downarrow}=2$.

We find always a non-magnetic (NM) solution to the mean-field equations. In addition, an antiferromagnetic (AFM) solution exists 
if $U_\alpha/t_\alpha>2$, that is always more stable than the NM solution whenever it exists. A ferromagnetic (FM) solution also 
exists if $U_\alpha/t_\alpha>4$. The FM solution is less stable than the AFM one, but more stable than the NM solution. 
Fig. \ref{fig:hubbard} is a graphical summary of these three solutions, where we draw the one-electron eigen-states,  
and write down the total energies and local magnetic moments. 

Figure \ref{fig:Utedge} (c) shows that the ratio $U_\alpha/t_\alpha$ increases monotonically as the ribbon length $M$ grows. 
We then define $M_{AFM}$ and $M_{FM}$ as the critical edge state lengths for which $U_\alpha/t_\alpha=2$ and $U_\alpha/t_\alpha=4$,
respectively.  
The expected behavior of a given edge state $p$  as a function of the ribbon length  $M$ can be summarized as follows,  
where we assume that $M_{c}<M_{AFM}$. For very short ribbons $M<M_{c}$, no edge state exists. 
Once $M>M_{c}$,  a NM edge state emerges. If $M$ grows beyond $M_{AFM}$, the edge state becomes AFM.  And if $M>M_{FM}$, 
both AFM and FM solutions can be found for the edge state, with the AFM solution being more stable in all cases.

\begin{figure}[tb!] 
\centering
\includegraphics[width=1.00\columnwidth]{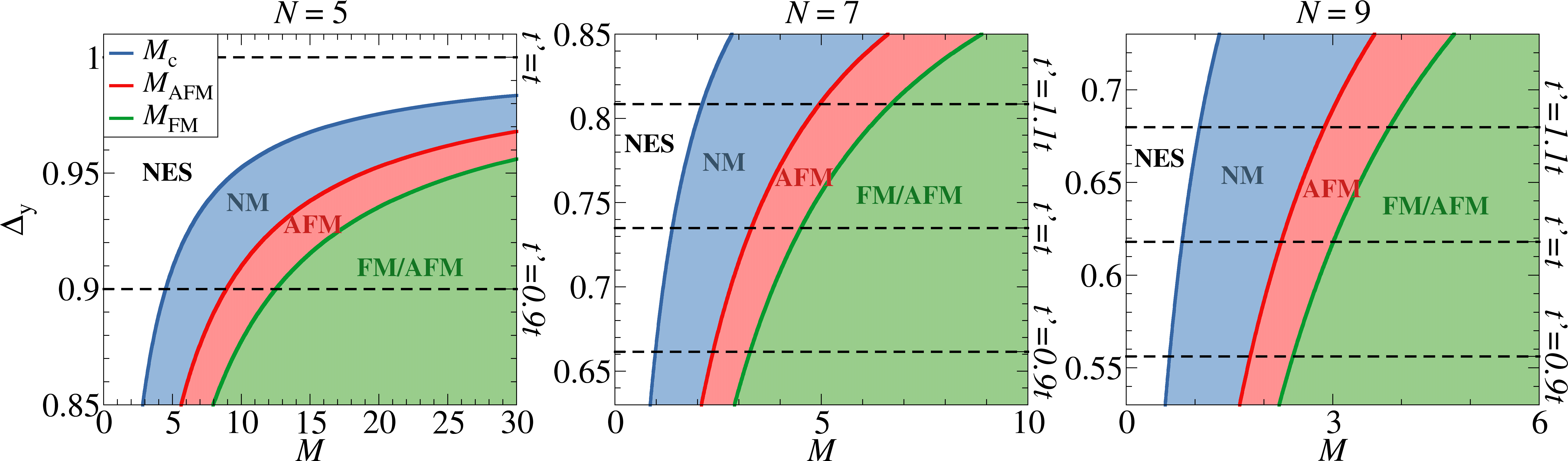}
\caption{$M_c$, $M_{AFM}$ and $M_{FM}$ as a function of $\Delta_\alpha$ and $M$ for ribbons with $N=5, 7,$ and $9$. Regions with no edge states (NES), 
and non-magnetic (NM), antiferromagnetic (AFM) and ferromagnetic (FM) solutions are marked by different colors. The scale in each of the three 
figures is different to help focusing on the different relevant ranges of $\Delta_\alpha$ and $M$ for each case. Dashed black lines indicate 
different reference values of $t'$.}
\label{fig:Mmag}
\end{figure}

We analyse now whether the three magnetic states can be realized in short-width ribbons that host a single edge state. 
Although at this point we do not know the exact values of the parameters that define the ribbon, we can make an educated 
guess that may shed some light on the expected behavior of the ribbons. We consider ribbons of $N=5,7,$ and $9$,
 and we estimate $t=U$. Then, for each ribbon we can calculate
$M_c$, $M_{AFM}$ and $M_{FM}$ as a function only of $\Delta_\alpha$ (that, for each value of $N$, only needs $\frac{t'}{t}$ to be defined).
We show our results in Fig. \ref{fig:Mmag}, where we focus especially
in the $\Delta_\alpha$ region where $\frac{t'}{t}\in \left [ 0.9, 1.1 \right ]$. In all cases we find the 4 types of behavior, but both 
$N=7$ and $N=9$ ribbons reach $M_{FM}$ already for ribbons with a few unit cells. More interesting is what happens with $N=5$ ribbons. 
In this case, $M_c$, $M_{AFM}$ and $M_{FM}$ all become much larger, and we can expect to be able to distinguish a quite wide range of 
integer $M$ values within each regime.

\section{DFT simulations of finite-length GNRS} \label{Sec:dft}

The goal of this section is two-fold. We want to check in the first place whether our results and predictions above using a simple 
TB model agree with more realistic DFT simulation. Second, we wish to determine the $U$, $t$ and $t'$ parameters of
our model that reproduce the DFT simulations.

We have performed DFT simulations of finite graphene nanoribbons of widths $N=5, 7,$ and $9$ and different lengths 
from $M=2$ to $M=10$ or $30$, depending on the width. We have used for this task the code 
SIESTA.\cite{siesta,siesta2} 
The choice is based on the fact that the SIESTA code expands wave-functions into a variational basis
of atomic-like functions. Therefore the SIESTA Hamiltonian is already written in the TB language. Difficulties arise
however because (a) SIESTA's atomic-like functions are not orthogonal to each other; (b) SIESTA's basis includes usually 
multiple-$\zeta$ atomic functions at each atom, that have the same angular symmetry (e.g.: two or three $s$-wave-functions, etc.);
(c) atomic-like functions have a radius larger than several times the inter-atomic distance, so that hopping integrals exist to
several neighbor shells. We shall explain below our procedure to handle these difficulties and achieve an accurate mapping.
  
\subsection{Simulation details}
We have chosen the generalized gradient approximation (GGA) parametrized by Perdew, Burke 
and Ernzerhof (PBE) \cite{pbe} for the exchange and correlation potential.
The code SIESTA uses the pseudopotential method as implemented by Troullier and Martins,\cite{Troullier1993} where core electrons
are integrated out and valence electrons feels semi-local potentials. We have employed standard pseudopotential parameters for
both carbon and hydrogen atoms.
We have employed a double $\zeta$ polarized (DZP) basis set for the carbon atoms, that includes 2 
pseudo-atomic orbitals for each 2$s$ and 2$p$ atomic state, and a $p$-polarized (e.g.: a $d$) function; we have used a simpler 
double $\zeta$ basis set for H with 2 orbitals for its 1$s$ states.
We have used a real-space grid defined by a mesh cut-off of 250 Ry. We have also relaxed all atom positions in the nanoribbons 
simulated until all forces were smaller than 0.001 eV/\AA. We have employed our own MATLAB scripts to post-process the SIESTA 
Hamiltonian.

\subsection{Tight-Binding model accuracy and parameters }
We have searched for NM, AFM and FM DFT self-consistent solutions for each of the ribbons that we have simulated. 
We have found that all those ribbons have a NM solution while AFM and FM solutions only exist for ribbons larger than 
given critical lengths. These facts fully agree with the TB and Hubbard dimer model predictions. 
We have taken advantage of  the fact that DFT is in effect a mean-field method. This means that we can use the 
Kohn-Sham (KS) eigen-energies to perform estimates and make comparisons with the eigen-energies of both the TB and 
the Hubbard dimer models, by using the equations in Fig. \ref{fig:hubbard}.

\begin{figure}[t!] 
\centering
\includegraphics[width=1.00\columnwidth]{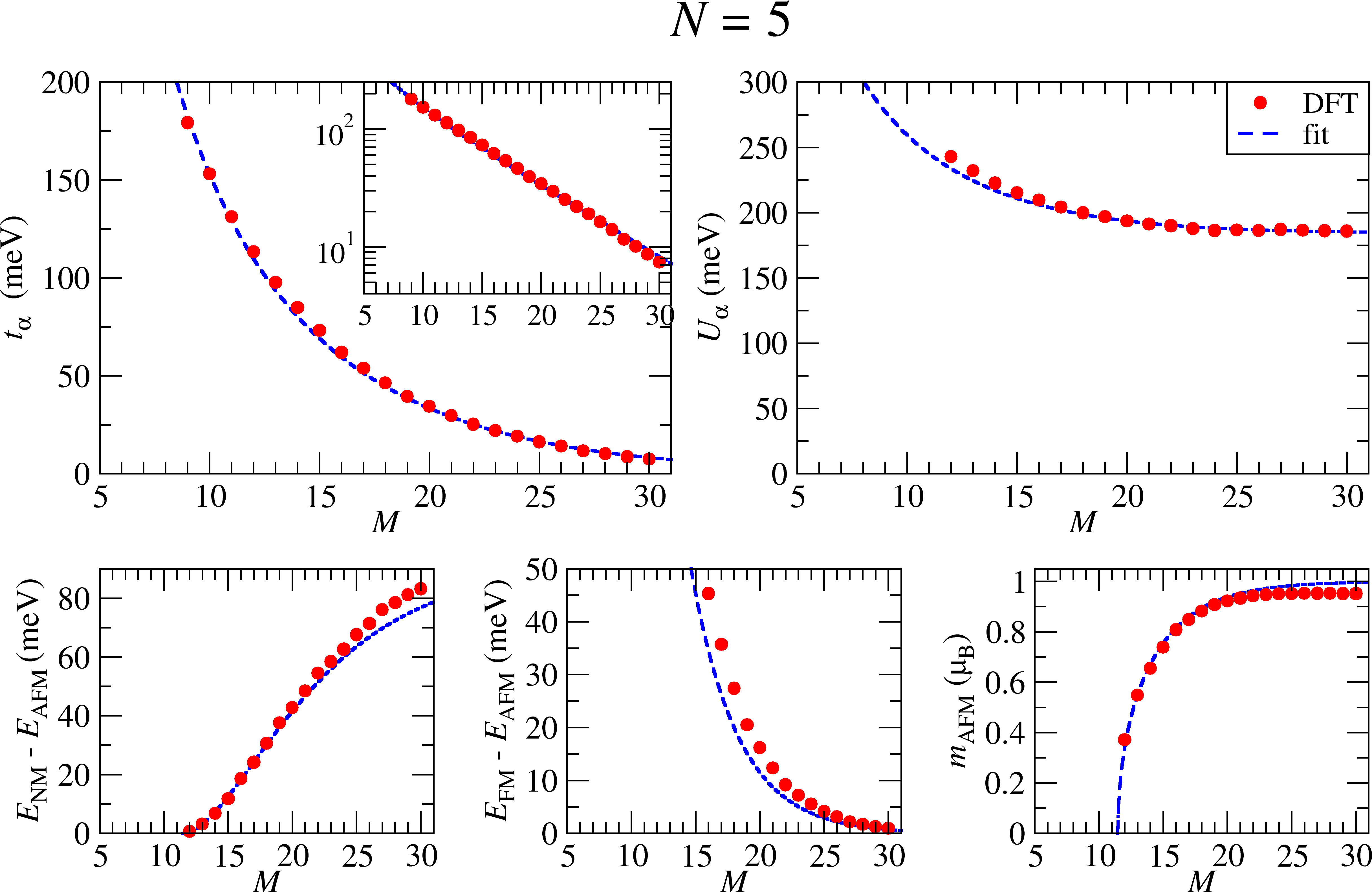}
\caption{(a) Effective hopping $t_\alpha$, (b) effective Hubbard-$U$ $U_\alpha$, (c) total energy difference between the NM and AFM solutions 
$E_{NM}-E_{AFM}$, (d) total energy difference between the FM and AFM solutions $E_{FM}-E_{AFM}$ and (e) magnetization $m$ as a 
function of the ribbon length $M$ for $N=5$ AGNR. The red dots correspond to our DFT simulations, while the blue dashed lines are 
the results of our TB model where we have fitted them in the upper two panels to obtain the optimal values of $U$ and $t$. The 
results in the lower three figures are parameter-free.
}
\label{fig:N5fit}
\end{figure}

\begin{table*}
\caption{Hubbard dimer model parameters $t$, $t'$ and $U$, obtained by fitting the TB estimates for $t_\alpha$ and $U_\alpha$ to 
the corresponding DFT results. The Table also includes the model $\Delta_\alpha$, $M_c$, $M_{AFM}$ and $M_{FM}$, and their DFT counterparts 
($M_{c}^{DFT}$, $M_{AFM}^{DFT}$ and $M_{FM}^{DFT}$).}
\begin{ruledtabular}
\begin{tabular}{lcccccccccc}
$N$ & $t$ (meV) & $t'$ (meV) & $U$ (meV) & $\Delta_\alpha$ & $M_{c}$ & $M_{AFM}$ & $M_{FM}$ & $M_{c}^{DFT}$ & $M_{AFM}^{DFT}$ & $M_{FM}^{DFT}$ \\
\colrule
5 & 4027.4 & 3758.7 & 5348.1 & 0.933 & 6.99 & 11.46 & 17.14 & 9 & 12 & 16 \\
7 & 3881.7 & 3422.8 & 3872.6 & 0.675 & 1.04 & 2.51 & 3.46 & 2 & 3 & 3 \\
9 & 5322.7 & 4089.3 & 3480.7 & 0.452 & 0.45 & 1.68 & 2.15 & 2 & 2 & 2
\end{tabular}
\end{ruledtabular}
\label{table:dftfit}
\end{table*}

First, we note that the eigen-energy of any bulk/edge state must lie inside the band/gap of the corresponding infinite-length 
ribbon. We can therefore simply look into the NM DFT solutions to establish the critical length $M_{c}^{DFT}$ as the
length in which in-gap states nucleate for the first time. Second, we can extract the effective hopping between DFT edge states 
$t_\alpha^{DFT}$ from the NM edge states KS eigen-energies (see the top panel in Fig. \ref{fig:hubbard}): 
\begin{equation}
t_\alpha^{DFT}=\frac{\varepsilon_{LUMO}^{NM}-\varepsilon_{HOMO}^{NM}}{2}
\end{equation}
Third, we can extract the Hubbard-$U$ interaction between DFT edge states $U_\alpha^{DFT}$ from the AFM edge states KS eigen-energies: 
\begin{equation}
U_\alpha^{DFT}=\varepsilon_{LUMO}^{AFM}-\varepsilon_{HOMO}^{AFM}
\end{equation}
We can then extract the TB parameters $t$, $t'$ and $U$ by fitting $t_\alpha$ in Eq. (\ref{eq:tp}) to $t_\alpha^{DFT}$ and 
$U_\alpha$ in Eq. (\ref{eq:Up}) to $U_\alpha^{DFT}$.

We show the results of this fitting procedure for $t_\alpha$ and $U_\alpha$, for $N=5$ ribbons, in the top two panels of Fig. \ref{fig:N5fit}.
We then write down in Table \ref{table:dftfit} the fitted values of $t$, $t'$ and $U$. We estimate now $\Delta_\alpha$, $M_{c}$, $M_{AFM}$ and
$M_{FM}$ from these fitted parameters, and compare them with the DFT values, that are also shown in Table \ref{table:dftfit}.
We stress that the two panels and the values of the critical lengths show that both model and DFT simulations agree truly well.
The high quality of the mapping can be further tested by looking into more complex magnitudes. We have chosen here the energy
differences between different magnetic solutions $E_{NM}-E_{AFM}$ and $E_{FM}-E_{AFM}$, as well as the magnetic moment of the 
AFM solution. The bottom panels in Fig. \ref{fig:N5fit} shed more weight on the quality of the mapping. We have chosen $N=5$ ribbons
for the present discussion because they have the highest potential for experimental testing of our predictions. The results for
$N=7$ and $9$ ribbons is qualitatively similar and therefore relegated to Appendix B.

Table \ref{table:dftfit} indicates a possible significant trouble for the validity of our results, since the fitted $t$ value
of about 4 to 5 eV is much larger than the universally accepted value for bulk graphene of about 2.7 eV.\cite{guinea} This discrepancy 
has prompted us to perform a deeper analysis of the DFT Hamiltonian.

\subsection{DFT Hamiltonian downsizing}
We devote this section to trim the SIESTA DFT Hamiltonian gradually from the initial full-basis form 
$H^{full}$ down to the simple TB expression given in Eq. (1).

Our first step is to reduce the basis set and leave only the $2 p_z$ carbon orbitals. This is equivalent to picking the 
Hamiltonian box containing only matrix elements among $2 p_z$ orbitals. We call the resulting Hamiltonian $H^{DZ}$ 
because each atom contains two $p_z$ orbitals. The drastic reduction of the Hamiltonian is justified by the fact that 
the lowest-lying valence and conduction bands of graphene have $2 p_z$ flavor to a very large extent.

The second step consists of reducing the basis from two $2 p_z$ orbitals per carbon atom to a single one. This is accomplished 
by making use of the variational principle and integrating out the unwanted high-energy degrees of freedom. The single 
remaining $p_z$ orbital is defined by the linear combination of the 2 original $p_z$ orbitals that minimizes the energy 
of the HOMO and LUMO states. We denote the resulting Hamiltonian $H^{SZ}$

SIESTA orbitals are non-orthogonal to each other, and so are the orbitals of the single-$\zeta$ basis defined in the previous paragraph.
We therefore compute the overlap matrix $S^{SZ}$ and orthogonalize the basis. The resulting Hamiltonian
$H^{SZ,orth}$ is already rather similar to the Hamiltonian in Eq. (1). There remain however three differences: first,
$H^{SZ,orth}$ has non-zero hopping integrals to first, second and third nearest neighbors, that we denote by 
$-t_1, t_2$ and $-t_3$, respectively; second, non-zero on-site energies $\varepsilon_0$ appear; third, both on-site energies
and hopping integrals are non-uniform across the ribbon.  We define $t_1$ and $t_3$ with a negative sign in front of them so that all
numbers are real positive. 

\begin{figure}[ht!] 
\centering
\includegraphics[width=1.00\columnwidth]{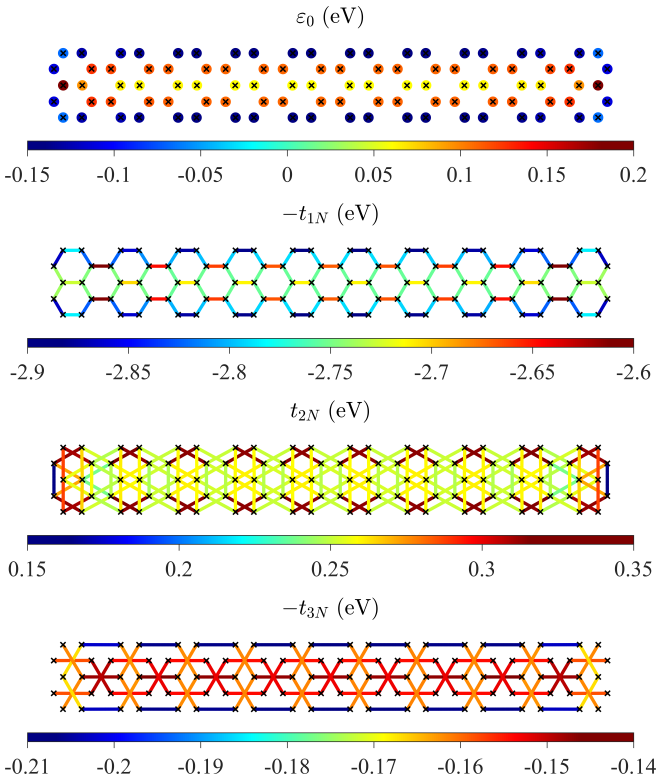}
\caption{Spatial representation of the different on-site energies $\varepsilon_0$ (referred to their average value) and 
first-, second- and thid-neighbor hopping integrals, $-t_1$, $t_2$ and $-t_3$ respectively, for a $N=5$, $M=10$ AGNR. 
Black crosses indicate the carbon atom positions.}
\label{fig:siesta}
\end{figure}

\begin{figure}[ht!] 
\centering
\includegraphics[width=1.00\columnwidth]{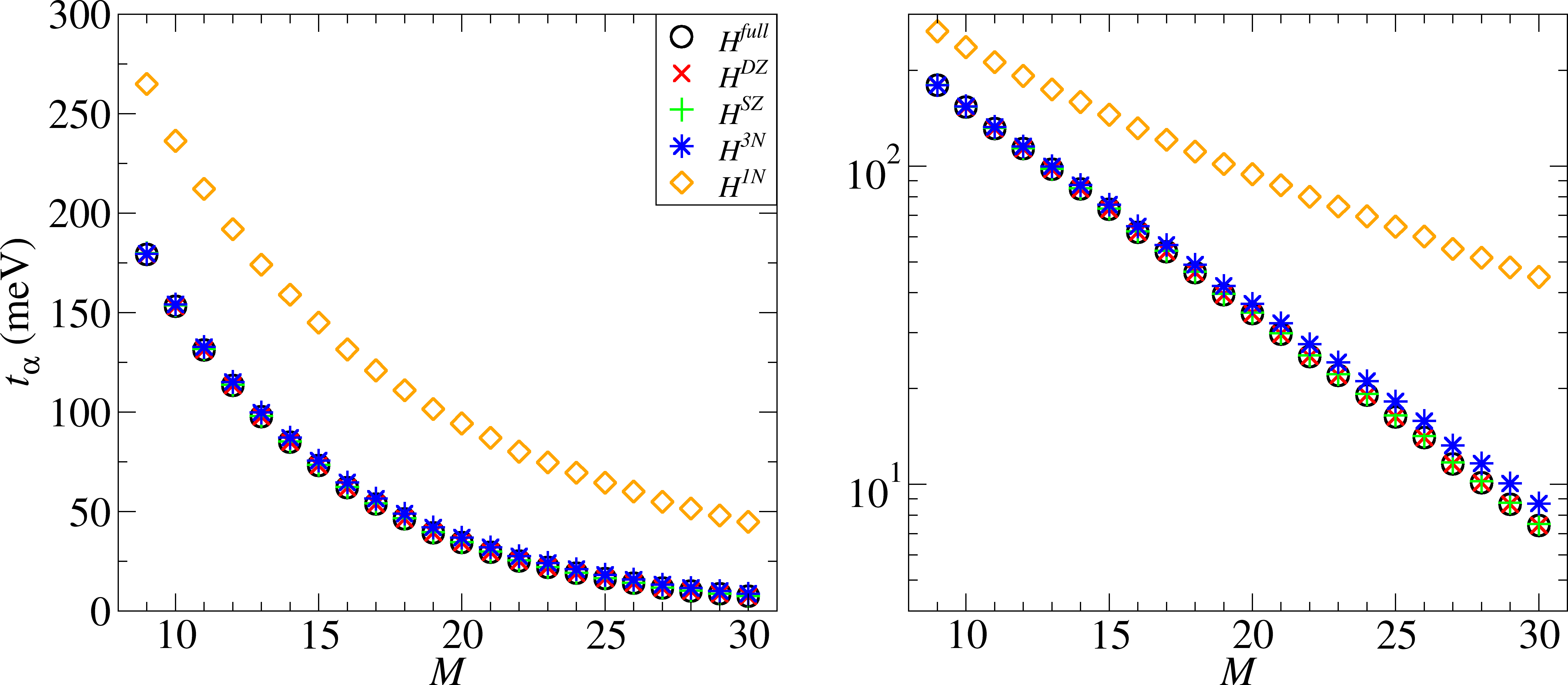}
\caption{$t_\alpha$ as a function of $M$ for a $N=5$ AGNR. Different curves refer to different versions of the DFT Hamiltonian:
$H^{full}$, $H^{DZ}$, $H^{SZ}$, $H^{3N}$ and $H^{1N}$.}
\label{fig:t_siesta}
\end{figure}

We show in Fig. \ref{fig:siesta}  the spatial distribution of on-site energies and hopping integrals for a $N=5$, $M=10$ ribbon 
to achieve further insight on their non-uniformities.  The figure shows that all values of $t_1$ fall in the range
(2.6, 2.9) eV in agreement with the accepted values of nearest neighbor hopping integrals in graphene.\cite{guinea}
We find that $\varepsilon_0\sim t_2\sim t_3$, and that the three are one order of magnitude smaller than $t_1$.
This later fact has prompted us to undertake two further trimmings on the Hamiltonian. The first consists of setting all on-site energies 
to zero, the resulting Hamiltonian being called $H^{3N}$. A second trimming consists of picking $H^{3N}$ and chopping off all $t_2$ and
$t_3$ hopping integral, whereby the resulting Hamiltonian $H^{1N}$ indeed conforms to Eq. (1).

We assess now the impact of each the above Hamiltonian reductions for a $N=5$ ribbon.  We show first $t_\alpha$ computed from the different
Hamiltonians as a function of the ribbon length in Fig.  \ref{fig:t_siesta}. We find that all of them deliver estimates for $t_\alpha$ in 
close agreement to the full DFT Hamiltonian. The single exception is $H^{1N}$, the one Hamiltonian that looks like Eq. (1). We then 
reach the conclusion that the simplest DFT-based Hamiltonian that reproduces the simulations is $H^{3N}$.

\begin{table}
\caption{First, second and third nearest neighboring hopping integrals in a infinite-length $N=5$ AGNR, as defined in 
Fig. \ref{fig:Z2} (a).}
\begin{ruledtabular}
\begin{tabular}{cccccc}
\multicolumn{2}{c}{$t_{1N}$ (meV)} & \multicolumn{2}{c}{$t_{2N}$ (meV)} & \multicolumn{2}{c}{$t_{3N}$ (meV)} \\
\colrule
$t_{1N}^{a}$ & 2713 & $t_{2N}^{a}$ & 259 &$t_{3N}^{a}$ & 149 \\
$t_{1N}^{b}$ & 2753 & $t_{2N}^{b}$ & 252 &$t_{3N}^{b}$ & 154 \\
$t_{1N}^{c}$ & 2694 & $t_{2N}^{c}$ & 312 &$t_{3N}^{c}$ & 161 \\
$t_{1N}^{d}$ & 2772 & &  &$t_{3N}^{d}$ & 208 \\
$t_{1N}^{e}$ & 2899 &  &  & &  \\
\end{tabular}
\end{ruledtabular}
\label{table:dftfit2}
\end{table}

\subsection{Parameter mapping}
Fig. \ref{fig:siesta} shows that the hopping integrals $t_i$ are mainly affected by their proximity to the edges, so that we should assess
whether those changes modify the topological protection and existence of edge states defined by the full Hamiltonian. To do so,
we define a new TB Hamiltonian for infinite-length $N=5$ ribbons $H^{TB}$ whose hopping integrals are defined graphically in Fig.
\ref{fig:Z2}, and are written down in Table \ref{table:dftfit2}. The hopping integrals $t_{1a}, t_{1c}$ and $t_{1e}$ correspond to
TB model $t$, while $t_{1b}$ and $t_{1d}$ correspond to $t'$. The table displays some apparent paradoxes because $t_{1a}, t_{1c}$ 
and $t_{1e}$ are not equal, and furthermore they are not really larger than $t_{1b}$ and $t_{1d}$, which is a requisite for the appearance
of an edge state for the $N=5$ ribbon within the TB model.

\subsection{$Z_2$ invariant}
\begin{figure*}[ht!] 
\centering
\includegraphics[width=1.00\textwidth]{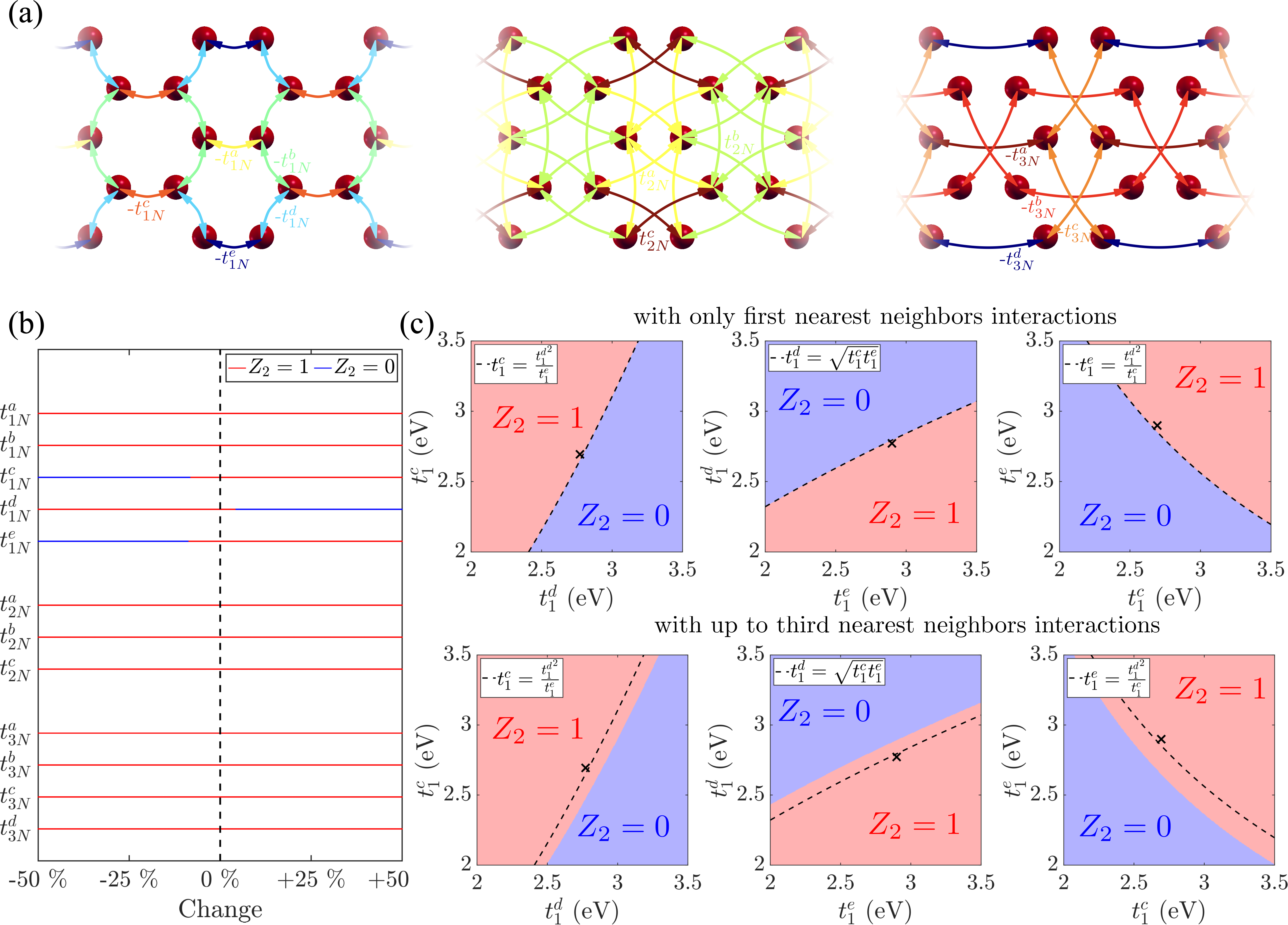}
\caption{
(a) Graphical definition of the different  hopping integrals $t_{1i}$, $t_{2i}$ and $t_{3i}$ in infinite-length $N=5$ ribbons. 
The color palette is consistent with the color scale shown in Fig. \ref{fig:siesta}.
(b) Values of $Z_2$ obtained from the bulk Hamiltonian defined in (a) and Table \ref{table:dftfit2}, when each of the hopping integrals is individually changed by a percentage of $\pm$ 50 \%.
(c) Values of $Z_2$ when $t_{1N}^{c}$, $t_{1N}^{d}$ and $t_{1N}^{e}$ are changed by pairs, considering only $t_{1N}$ interactions (upper panels) or $t_{1N}$, $t_{2N}$ and $t_{3N}$ interactions
(lower panels). The dashed lines indicate the values $t_{1N}^{d} = \sqrt{t_{1N}^{c} t_{1N}^{e}}$  ($t'=t$) The crosses indicate the reference values of each couple of $t_{1N}$.}
\label{fig:Z2}
\end{figure*}

We have computed the $Z_2$ invariant using $H^{TB}$, and have found that $Z_2=1$ as expected, hence confirming the presence of 
topologically protected edge states. We modify now each of the different hopping integrals in the model at a time to identify which
of them affect most the $Z_2$ value. Our results, shown in Fig. \ref{fig:Z2} (b), demonstrate that changes in any $t_{2}$, $t_{3}$, or 
 $t_{1a}, t_{1b}$ do not modify $Z_2$, while small variations in $t_{1c}, t_{1d}$ or $t_{1e}$ do, and kill the edge states.

For a moment, let's just focus on a first neighbor Hamiltonian. In this case, our model indicates that for $N=5$, inside the relevant region of the reciprocal space to find edge states $k_\alpha=\frac{\pi}{3}$,
and the coefficients $E_{y}(k_y)$ defined in equation (\ref{eq:crjbulk2}) vanish in the central row of the ribbon $R_y=3$. This condition, fixed by the boundary conditions
in the Y direction, is maintained even if we change the different values of $t_{1N}$, as far as the axial symmetry around the axis defined by the central row of C atoms
is conserved. Then, any interaction with the central C-atoms of the ribbon, that is, $t_{1N}^{a}$ and $t_{1N}^{b}$, becomes irrelevant for the properties of the edge states;
and the edge states of the ribbon are exactly those of a SSH-like chain with $t$ and $t'$ hopping integrals, formed by the 2 upper or 2 lower C chains of the ribbon structure (as it is clear from 
the value of $f({\bf k})=t+t' e^{i k_x}$). 

We show in Fig. \ref{fig:Z2} (c) the value of $Z_2$ of the ribbon as we modify $t_{1N}^{c}$, $t_{1N}^{d}$ and $t_{1N}^{e}$ in pairs, considering only the $t_{1N}$ interactions (upper panels),
or all the interactions shown in (a) (lower panels). With only the $t_{1N}$ interactions, we can make a correspondence $t' \leftrightarrow t_{1N}^{d}$,  $t \leftrightarrow \sqrt{t_{1N}^{c} t_{1N}^{e}}$.
Similar relations between our simplified $t,t'$ parameters and the $t_{1N}$ values of the real ribbon are expected for ribbons of other widths. 
Then, we obtain that the transition between $Z_2=0$ and $Z_2=1$ occurs exactly at $t=t'$, as expected in our model.
Cao {\it et al}\cite{Cao2017} indicate that a distortion at the edges leading to a stronger hopping between the edge atoms ($t_{1N}^{e}$ in our calculations) is enough to open a GAP in the band
structure of these ribbons and obtain $Z_2=1$ for $N=5$. This agrees with our results, where the strongest value of $t_{1N}$ is indeed $t_{1N}^{e}$, and is crucial to fulfill the $t'<t$ condition.
However, we go beyond this edge-distorted model, as we consider the effect of changing any of the $t_{1N}$ parameters.

With the values of Table \ref{table:dftfit2}, $Z_2=1$ but $\Delta_y=0.997$ and
$M_c \simeq 187$, that explains why no edge states are shown in Fig. \ref{fig:t_siesta} for $H^{1N}$.
Including $t_{2N}$ and $t_{3N}$ interactions changes the results, increasing
the region where $Z_2=1$. Although several factors affect to this change, the most important is the inclusion $t_{3N}^{b}$ and $t_{3N}^{d}$ that modify the SSH-like chain formed by 2 C chains.
If we include in our model an average $t_{3N}=\sqrt{t_{3N}^{b} t_{3N}^{d}}$, $f(\bf{k})$ in the SSH-like chain becomes:

\begin{equation}
\label{eq:fk_3nn}
\begin{aligned} 
f({\bf k})= & t+t' e^{ik_x}+t_{3N} e^{i 2 k_x} = \\ = & (t-t_{3N}) + (t'+2t_{3N} \cos{ \left ( k_x \right )}) e^{i k_x}
\end{aligned}
\end{equation}

In this case, when $k_x \rightarrow \pi$, $\theta_{\bf k} \rightarrow 0$ if $t'-t_{3N}<t$, and the condition to obtain edge states becomes less restrictive, in agreement to what is shown in Fig. \ref{fig:Z2} (c).
We can make a rough estimation of the equivalent $\Delta_\alpha = \frac{t'-t_{3N}}{t} = 0.927$, in good agreement with the result of $\Delta_\alpha = 0.933$ obtained from our fitting of $t$ and $t'$. Therefore,
we can assume that the obtained value of $\Delta_y$ in our fitted TB model, that is the main responsible of the behavior of the edge states in the ribbon, is correct, but it is obtained at the cost
of getting unrealistic values of $t$ and $t'$ that take care of the effects of interactions between other neighbors and of the differences in the hopping integrals as we move closer to the edges. 


\section{Conclusions} \label{Sec:conclusions}
We have presented a full analytical solution of the TB model of finite-length AGNRs, that we have also called {\it rectangulenes}.
We have indeed shown that the above problem can be separated as the product of a one-dimensional finite-length mono-atomic chain times
a one-dimensional finite-length dimerized chain. We have written down the explicit expressions for the quantum numbers, the eigen-functions 
and the eigen-energies. We have found that finite-length armchair ribbons witness a cascade of magnetic transitions as a function of the
ribbons length. We have found ample room for experimental testing of the prediction in $N=5$ AGNRs. 

We have also performed DFT simulations of $N=5, 7$ and $9$ ribbons where the above TB-based estimates are confirmed. We have
then performed a mapping between the TB and the DFT Hamiltonian to check the robustness of the predictions and determine the
model parameters.  


\section{Acknowledgments}
The research carried out in this article was funded by project PGC2018-094783 (MCIU/AEI/FEDER, EU) and by Asturias FICYT under grant AYUD/2021/51185 with the support of FEDER funds.
G. R. received a GEFES scholarship.


\appendix

\section{Open boundary conditions in TB chains} \label{Appendix:obc}

\begin{figure}[ht!] 
\centering
\includegraphics[width=1.00\columnwidth]{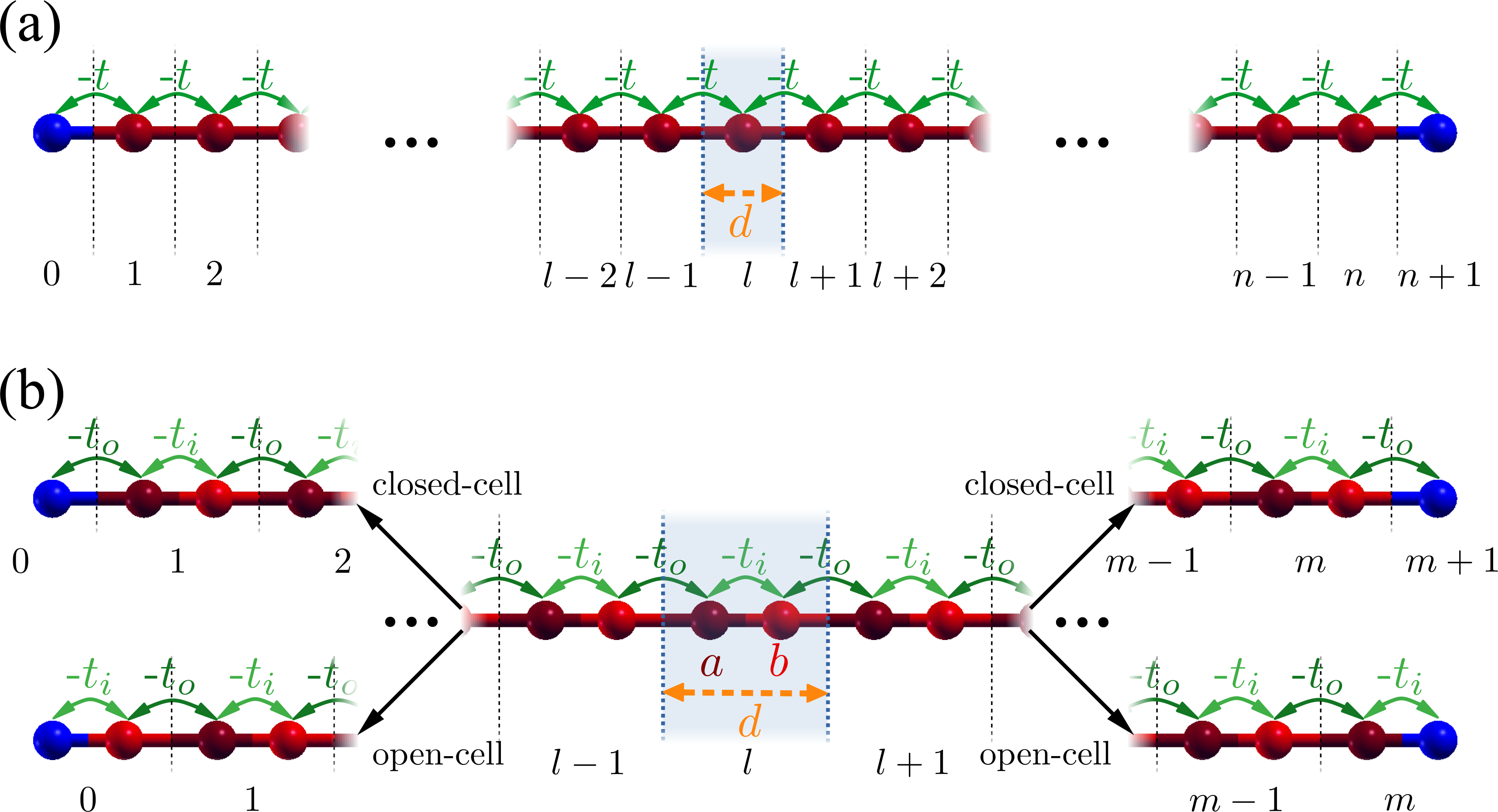}
\caption{Structure of (a) a monoatomic chain of $n$ atoms and (b) a dimerized chain (SSH) of $2m$ atoms, with open boundary conditions. Unit cells of lattice constant $d$ are separated by dotted lines
and labeled in black, with a central, $l$ cell shaded in blue. Red spheres represent atoms belonging to the chain, with a light and dark red used in the dimerized chain to differentiate between the two types of site, labeled
$a$ and $b$. Blue spheres represent $fake$ atoms used to define the open boundary conditions. Different hopping integrals are shown in different shades of green. The closed-cell and open-cell structures at the
edges of the SSH chain are also represented.}
\label{fig:Astructures}
\end{figure}

In this appendix we show the analytical solution of the TB Hamiltonian of a monoatomic chain (Fig. \ref{fig:Astructures} (a)) and of a dimerized chain (Fig. \ref{fig:Astructures} (b)), also known as the SSH model,\cite{Su1979} with
open boundary conditions.

The solution of the monoatomic chain is quite straightforward. We consider a chain of $n$ sites (where we use lower case letters to avoid confusion with the definition of the graphene
ribbon structure in the main text), with all on-site energies shifted to zero and first neighbors interaction of value $-t$. In the basis of the orbitals located on each cell $l$, labeled $ |l \rangle$, any wave-function can be
described from a set of coefficients $C_l$ as:

\begin{equation}
| \Psi \rangle = \sum_{l} C_l |l \rangle
\end{equation}

In particular, a Block wave-function of the system $| u_k \rangle$ can be written as:

\begin{equation}
| u_k \rangle  = \sum_{l} e^{ikl} |l \rangle
\end{equation}

\noindent
where $k$ is measured in units of the inverse of the lattice constant, $d^{-1}$. The expression of the energy for $| u_k \rangle$, $\varepsilon_k = -2t \cos{\left ( k \right )}$, leads to a degeneracy
$\varepsilon_{k}=\varepsilon_{-k}$. Therefore, we write the following trial wave-function, of energy $\varepsilon_{k}$:

\begin{equation}
\label{eq:APblochcomb}
| \Psi \rangle = A_1 | u_k \rangle + A_2 | u_{-k} \rangle
\end{equation}

\noindent
to try to fulfill the open boundary conditions, consisting in:

\begin{equation}
C_0 = C_{n+1} = 0
\end{equation}

We obtain the following solution:

\begin{equation}
\label{eq:Akmono}
C_l = A \sin{\left ( kl \right )}; k=\frac{\alpha \pi}{n+1}; \alpha=1...n
\end{equation}

\noindent
where $A$ is just a normalization constant.

We define the dimerized chain (Fig. \ref{fig:Astructures} (b)) as follows. Each unit cell $l$ contains 2 orbitals $a$ and $b$, so we write $| l, a \rangle$, $| l, b \rangle$ to identify
our basis. Those can be gathered in a single vector for each cell as:

\begin{equation}
|l\, \rangle=\left(\begin{matrix} |l,a \rangle\\ |l,b \rangle\end{matrix} \right )
\end{equation}

All on-site energies are shifted to zero, and each orbital of type $a$ ($b$) interacts only with its neighbors of type $b$ ($a$) with an interaction labeled
$-t_i$ or $-t_o$ depending on if it occurs within the same unit cell of between neighboring cells. In this basis, any wave-function can be written as:

\begin{equation}
| \Psi \rangle = \sum_{l} C_l^{\top} | l \rangle
\end{equation}

\noindent
while Bloch wave-functions $| u_k \rangle$ verify:

\begin{equation}
C_l=\left(\begin{matrix} \,c_l^a \\ \,c_l^b\end{matrix}\right)=e^{ik\,l}\,C_{k} = e^{ik\,l}\,\left(\begin{matrix} \,c_{k}^a \\ \,c_{k}^b\end{matrix}\right)
\end{equation}

\noindent
where the coefficients $c_k^a$ and $c_k^b$ have to be obtained from the diagonalization of a $2 \times 2$
effective Hamiltonian:

\begin{equation}
\label{eq:APhbulk}
\begin{aligned} 
H &= \left ( \begin{matrix}
0 & -t_i -t_o e^{-ik} \\
 -t_i -t_o e^{ik} & 0 \end{matrix}  \right ) = \\ & = 
 \left ( \begin{matrix} 0 & -t_i \left (  1 + \Delta e^{-ik} \right ) \\ -t_i \left (  1 + \Delta e^{ik} \right ) & 0 \end{matrix} \right ) = \\ & =
 \left ( \begin{matrix} 0 & -f^{*}(k) \\ -f(k) & 0 \end{matrix} \right ) = 
- \left | f(\bf{k}) \right | \left ( \begin{matrix} 0 & e^{-i\theta_{k}}\\ e^{i\theta_{k}} & 0 \end{matrix} \right )
\end{aligned}
\end{equation}

\noindent
where we defined $\Delta = \frac{t_o}{t_i}$, $f(k) = t_i +t_o e^{ik}$ and $\theta_k$ as the polar angle of the complex number $f(k)$. The Bloch
wave-functions are then described by:

\begin{equation}
c_k^a=1; c_k^b=\tau e^{i\theta_k}; (\tau = \pm)
\end{equation}

\noindent
with energy:

\begin{equation}
\varepsilon_k =- \tau \left| f(k) \right| = - \tau t_i \sqrt{1+\Delta^2 + 2\Delta \cos{\left ( k \right )}}
\end{equation}

We now focus on the open boundary conditions for a SSH chain of 2$m$ atoms. Like for the monoatomic chain, $\varepsilon_k = \varepsilon_{-k}$
and therefore we use the same linear combination of Bloch wave-funtions of equation (\ref{eq:APblochcomb}) as trial wave-functions. Two different cases
can be considered (Fig. \ref{fig:Astructures} (b)). If the chain contains only complete unit cells, we call this chain a closed-cell SSH chain. If the cells at the edges contain only
one atom belonging to the chain, we call this an open-cell SSH chain. It is clear that we can transform one system into the other by exchanging the labels $t_i$ and $t_o$.
Therefore, we solve explicitly the closed-cell case, and at the end we do the needed transformations to obtain the solution of the open-cell case, which is
relevant in the context of graphene ribbons.

The open boundary conditions at one edge define the general shape of the wave-function:

\begin{equation}
\begin{aligned} 
c_0^b = 0 \Rightarrow & c_l^a = A \sin{ \left ( kl - \theta_k \right ) }  \\ & c_l^b = \tau A \sin{\left ( kl \right )}
\end{aligned}
\end{equation}

\noindent
where $A$ is a normalization constant. The conditions at the other edge determine the possible values of $k$:

\begin{equation}
c_{m+1}^a=0\Rightarrow \sin{\left ( k \left ( m+1 \right ) - \theta_k \right )} = 0;
\end{equation}

\begin{equation}
\label{eq:APgk}
g (k) := k(m+1) - \theta_k = \beta \pi; \beta=1...m 
\end{equation}

This relation allows us to rewrite the coefficients $c_l^a$ as:

\begin{equation}
c_l^a = A (-1)^{p+1} \sin{\left( k \left ( m+1-l \right ) \right )}
\end{equation}

Equation (\ref{eq:APgk}) must be solved numerically, under the restriction that $k \in (0,\pi)$, as both $k=0$ and $k=\pi$ lead to $c_l^a=c_l^b=0$ for any $l$.
All these real values of $k$ lead to states delocalized over all the chain, that is, bulk states. However, unlike what happens for an infinite chain or for a chain
with periodic boundary conditions, in the finite chain the loss of translational symmetry opens the door to the existence of states located close to the limits of the chain,
that is, edge states. These states can also be described with a wave-vector $k$, but with an imaginary part. Our objective now is to determine wether these
states exist in the chain or not.

\begin{figure}[ht!] 
\centering
\includegraphics[width=1.00\columnwidth]{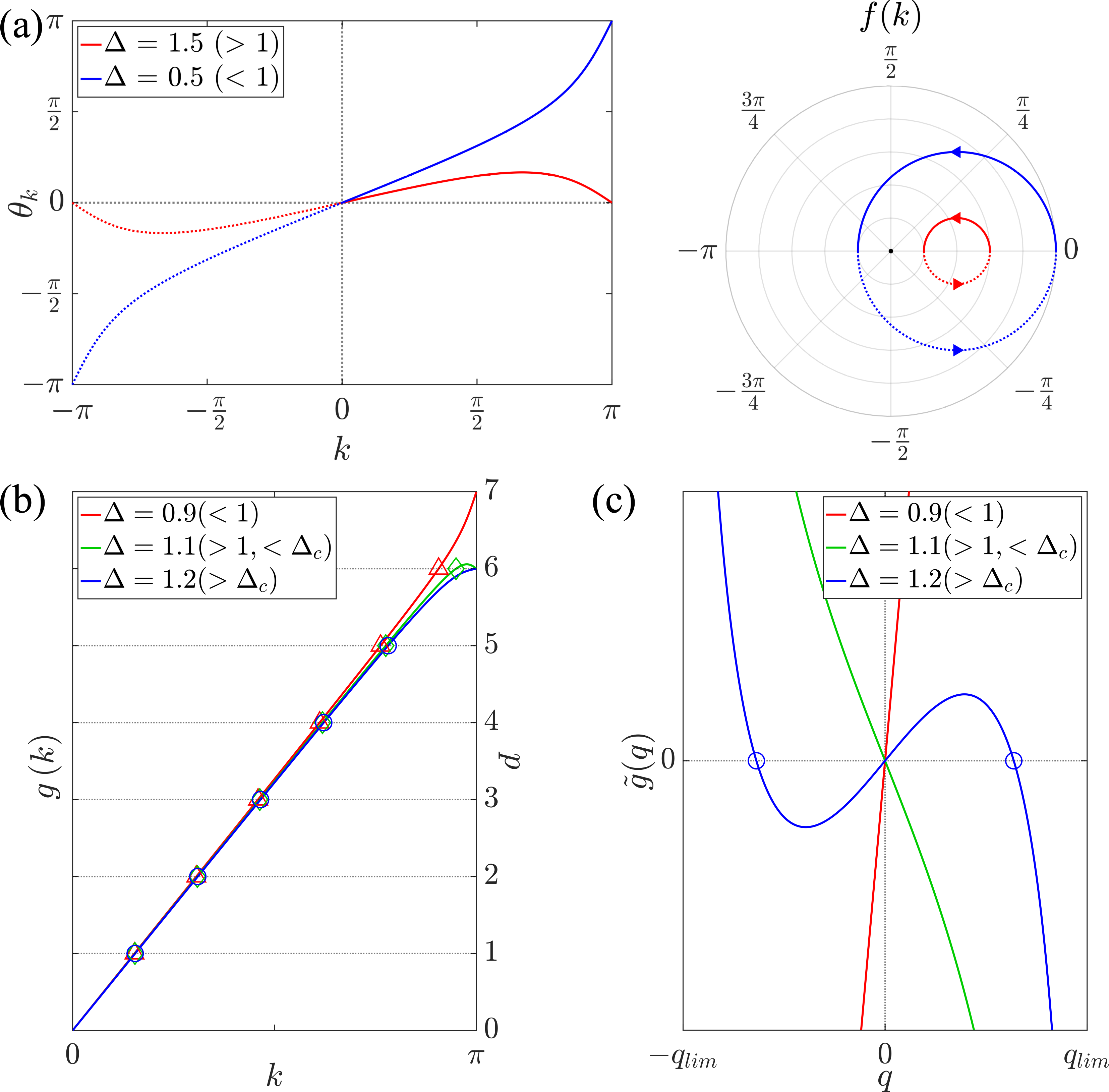}
\caption{(a) Representation of the evolution of $\theta_k$ and $f(k)$ as $k$ evolves through the first Brillouin zone for SSH chains with different
values of $\Delta$. Dotted lines are used in the range $k \in (-\pi, 0)$, while continuous lines are used in the range $k \in (0, \pi)$, relevant
for equation (\ref{eq:APgk}). (b) Representation of $g(k)$ for a SSH chain of $m=6$, with different representative values of $\Delta$. Different
symbols are used to identify the different real solutions of $k$ in equation (\ref{eq:APgk}). (c) Representation of $\tilde{g}(k)$ for the same chain
and $\Delta$ values of (b), with blue circles showing the imaginary part of the complex solutions for $\Delta > \Delta_c$.}
\label{fig:Amulti}
\end{figure}

The problem can be faced from the perspective of topology. The bulk-boundary correspondence establishes that we can define a topological invariant from the
bulk wave-functions, whose value determines the existence or not of edge states at the boundaries.\cite{laszlobook} This correspondence supposes a
closed-cell structure at the edges. In the case of a one dimensional system that can be described with a $2 \times 2$ Hamiltonian $H(k)$ in terms of the Pauli matrices $\sigma_x$ and
$\sigma_y$ from a two dimensional vector $\vec{d(k)}=(d_x(k),d_y(k))$ as:

\begin{equation}
H(k)=d_x(k)\sigma_x + d_y(k)\sigma_y = \vec{d} (k) \vec{\sigma}
\end{equation}

\noindent
the relevant topological invariant is the winding number $\nu$. $\nu$ is just the number of loops that $\vec{d}$ performs around the origin when
$k$ goes through the first Brillouin zone. Topology states that if $\nu=0$, all $k$ values are real and no edge states appear, while if $\nu=1$ there is a $k$
with an imaginary part that leads to a couple of edge states. Notice that, besides a global sign, $\vec{d}$ is just $f(k)$ in the XY instead of the complex plane.
Therefore, we can analyze $\nu$ by analyzing the evolution of $\theta_k$ as $k$ goes from $-\pi$ to $\pi$. 
Fig. \ref{fig:Amulti} (a) shows the evolution of $\theta_k$ through the first Brillouin zone, as well as the evolution of $f(k)$ in the polar plane. It is clear 
that $\nu=1$ ($\nu=0$) if $\Delta>1$ ($\Delta<1$).

Our chain of $m$ cells must contain $m$ values of $k$, whether real or complex. Looking into equation (\ref{eq:APgk}), if we had $\theta_k=0$ for all
values of $k$, $g(k)$ would be a straight line and the valid values of $k$ would be just those of a monoatomic chain of $n=m$ atoms, as shown in
equation (\ref{eq:Akmono}). As $\theta_k$ is a continuous function of $k$, the values of $k$ deviate from those of the monoatomic chain, but we know that
each time $g(k)$ crosses an integer value times $\pi$ in the range $k \in (0,\pi)$, a new real solution of $k$ arises. If $\Delta<1$, $\theta_k (0) = \theta_k (\pi) =0$,
the values of $g(0)$ and $g(\pi)$ do not change from those of the monoatomic chain, and therefore the existence of $m$ real values of $k$ is guaranteed
by the continuity of $g(k)$. If $\Delta>1$, however, $\theta_k(\pi)=\pi$ and $g(\pi)$ decreases a $\pi$-step from the monoatomic case. Therefore, continuity of $g(k)$
only guarantees the existence of $m-1$ real values of $k$.
This is exactly the result obtained from $\nu$. In other words, the winding number is just a measurement of the change of $\theta_k$ through the first Brillouin zone that reduces
the number of bulk states that can be guaranteed by continuity. However, this is not the whole story, as continuity of $g(k)$ only fixes a lower bound to the number
of bulk states, but it can not guarantee the existence of edge states. Looking at the behavior of $\theta_k$ as a function of $k$ for $\Delta>1$
(Fig. \ref{fig:Amulti} (a)), $\theta_k$ is a monotonous function of $k$ that increases first slowly, but finally fast as $k$ is close to $\pi$. Then, $g(k)$ can become a
decreasing function around $k=\pi$. In this case, an extra real value of $k$ appears and the system has no edge states, even although $\nu=1$.
This condition translates to:

\begin{equation}
\label{eq:APmck}
\left. \frac{d g\left ( k \right ) }{d k} \right |_{k=\pi}=m+1-\frac{\Delta}{\Delta-1}<0 \Rightarrow m < m_c = \frac{1}{\Delta-1}
\end{equation}

If the length of the chain $m$ is below a certain threshold $m_c$, we still have $m$ bulk values of $k$. Alternatively, for a fixed value of $m$, if $\Delta$ is below a critical
value $\Delta_c=\frac{m+1}{m}$, we also have $m$ bulk states. If this is not the case, we must find a complex
value of $k$. We show an example of the different possible behaviors of $g(k)$ in Fig. \ref{fig:Amulti} (b).

We search for complex values of $k$ by analytical continuation of $k$ in the limits of its validity range, $k=0-iq$ or $k=\pi-iq$. It can be demonstrated
that only the second case leads to a valid solution. The Hamiltonian of equation (\ref{eq:APhbulk}) then becomes:

\begin{equation}
\label{eq:ap_ssh_hedge}
\begin{aligned} 
H &= 
 \left ( \begin{matrix} 0 & -t_i \left (  1 - \Delta e^{-q} \right ) \\ -t_i \left (  1 - \Delta e^{q} \right ) & 0 \end{matrix} \right ) = \\ & =
f(q)  \left ( \begin{matrix} 0 & e^{-\theta_{q}}\\ e^{\theta_{q}} & 0 \end{matrix} \right )
\end{aligned}
\end{equation}

\noindent
where $f(q)$ is the geometric mean of the off-diagonal terms of the Hamiltonian (that is positive as $\Delta > 1$), and $\overline{\theta}_q = \frac{1}{2}\log{ \left ( \frac{1-\Delta e^q}{1-\Delta e^{-q}} \right ) }$
is introduced to mimic $\theta_k$ in equation (\ref{eq:APhbulk}). We require $q \in \left ( -q_{lim}, q_{lim} \right )$ to guarantee that $\overline{\theta}_q$ is real, with $q_{lim} = | \log{\left ( \Delta \right ) } |$.
The solutions of the Hamiltonian are then:

\begin{equation}
c_q^a=1; c_q^b=\tau e^{\overline{\theta}_q}; ( \tau = \pm )
\end{equation}

\noindent
with energy:

\begin{equation}
\varepsilon_q = \tau f(q) = \tau t_i \sqrt{1+\Delta^2 - 2\Delta \cosh{\left ( q \right )}}
\end{equation}

Once again, we have to apply the open boundary conditions, with the first one defining the general shape of the wave-function:

\begin{equation}
\label{eq:APcoefq1}
\begin{aligned}
c_0^b = 0 \Rightarrow & c_l^a = A (-1)^{l} \sinh{\left ( ql - \overline{\theta}_q \right ) } \\ & c_l^b = \tau A (-1)^{l}  \sinh{\left ( ql \right )}
\end{aligned}
\end{equation}

\noindent
where $A$ is a normalization constant. The conditions at the other edge determines the possible values of $q$:

\begin{equation}
\label{eq:APcoefq2}
c_{m+1}^a=0\Rightarrow \sinh{\left ( q \left ( m+1 \right ) - \overline{\theta}_q \right )} = 0;
\end{equation}

\begin{equation}
\label{eq:APgq}
\tilde{g} (q) := q(m+1) - \overline{\theta}_q = 0
\end{equation}

This relation allows us to rewrite the coefficients $c_l^a$ as:

\begin{equation}
\label{eq:APcoefq3}
c_l^a = A (-1)^{l+1} \sinh{\left ( q \left ( m+1-l \right ) \right )}
\end{equation}

Condition (\ref{eq:APgq}) is always satisfied for $q=0$, but this leads to the invalid, real solution $k=\pi$. Other possible
values of $q$ must be obtained numerically, but we can determine if these solutions exist by analyzing the behavior
of the function $\tilde{g}(q)$ (see Fig. \ref{fig:Amulti} (c)). For $\Delta<1$ we only find $\tilde{g}(0)=0$.
For $\Delta>1$ this function is continuous inside the defined range of $q$, odd, and
$\tilde{g}\left ( q \to \pm q_{lim}  \right )=\mp \infty $. Then, there are other two solutions of $\tilde{g}(q)=0$, of value $\pm q$, if:

\begin{equation}
\label{eq:APmcq}
\left. \frac{d \tilde{g}\left ( q \right ) }{d q} \right |_{q=0}=m+1-\frac{\Delta}{\Delta-1}>0 \Rightarrow m > m_c (\Delta > \Delta_c)
\end{equation}

Notice that solutions of value of $\pm q$ lead to the same coefficients of the wave-function in equations (\ref{eq:APcoefq1}) and (\ref{eq:APcoefq3}), up to a sign.
Therefore, it is enough to consider the solution with $q>0$. Results of equations (\ref{eq:APmck}) and (\ref{eq:APmcq}) 
are consistent. For a given chain defined by $\Delta$ and $m$, if $\Delta <1$, or $\Delta>1$ but $m<m_c$ (equivalent to $\Delta<\Delta_c$), the chain
presents $m$ real values of $k$ leading to $2m$ bulk solutions. If $\Delta>1$ and $m>m_c$ (equivalent to $\Delta>\Delta_c$), the chain contains
$m-1$ real values of $k$ to define $2m-2$ bulk states, but also a complex value of $k=\pi-iq$, leading to 2 localized
edge states.

The value of $q$ indicates the level of localization of the edge states, as $q^{-1}$ is a measurement of the penetration depth of the state
in units of $d$. The exact value of $q$ for a given value of $\Delta$ and $m$ must be obtained numerically solving equation 
(\ref{eq:APgq}), or any of the following, equivalent equations:

\begin{equation}
\tanh{ \left ( qm \right ) }=\frac{\sinh{ \left ( q \right ) }}{\Delta-\cosh{\left ( q \right )}}
\end{equation}

\begin{equation}
\Delta \sinh{ \left ( qm \right ) }= \sinh{\left ( q \left ( m+1 \right ) \right )}
\end{equation}
 
We can obtain an approximated value of $q$ if it is close to $q_{lim}$ with the following expression:

\begin{equation}
\label{eq:APqaprox}
q=q_{lim}-\frac{ \Delta^ 2-1}{1+\Delta^{2m+2}-2m\left ( \Delta^ 2-1 \right )}
\end{equation}

Alternatively, we propose the following iterative solution that, starting at $q_0$=$q_{lim}$, converges quickly to the exact value of $q$:

\begin{equation}
C_i=\tanh{\left ( mq_{i-1} \right )}
\end{equation}

\begin{equation}
q_i=\log{ \left ( \frac{\Delta C_i}{1+C_i} +\sqrt{\left ( \frac{\Delta C_i}{ 1+C_i } \right )^ 2+\frac{1-C_i}{1+C_i}}  \right ) }
\end{equation}

\begin{figure}[ht!] 
\centering
\includegraphics[width=1.00\columnwidth]{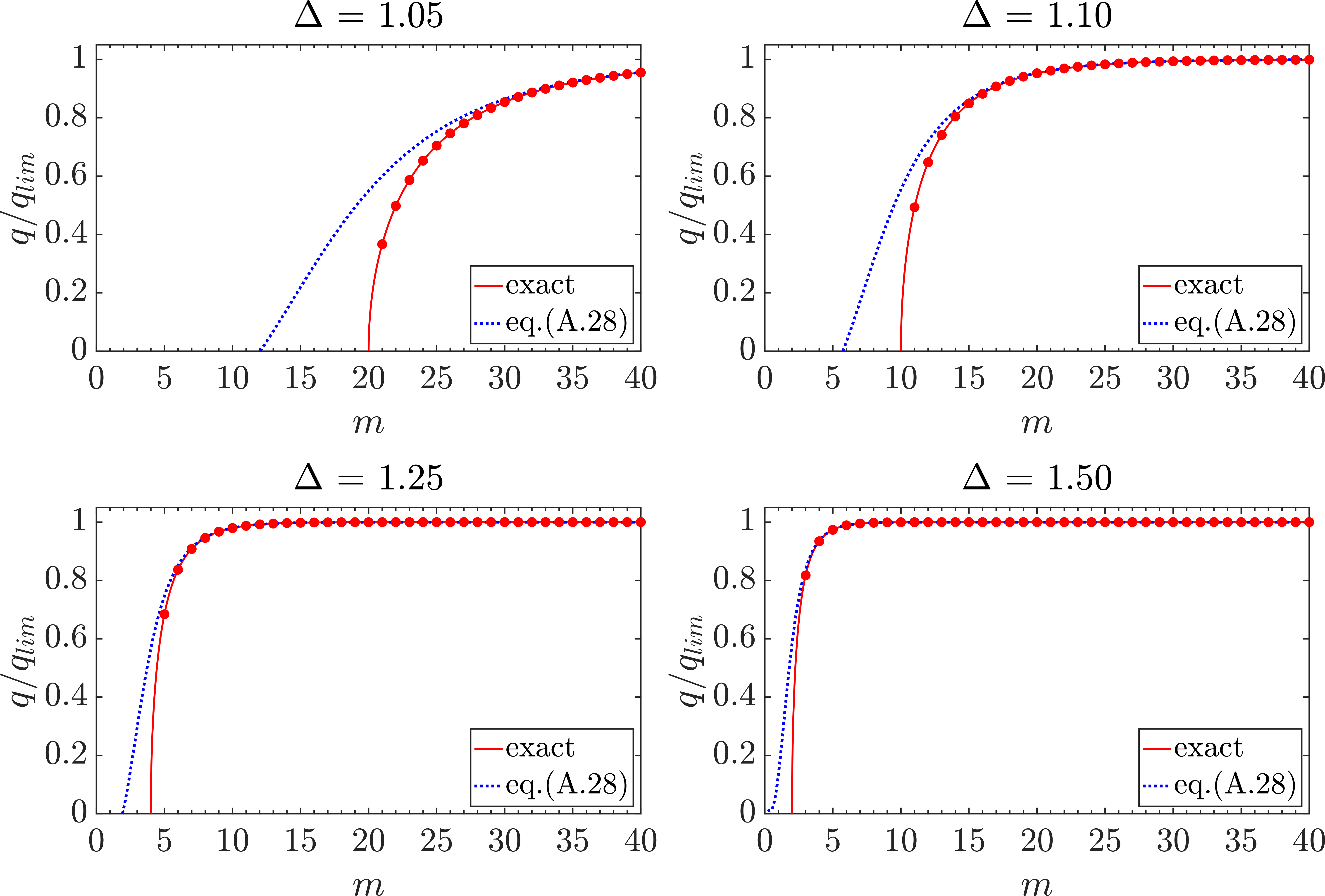}
\caption{Values of $q$ normalized by $q_{lim} = | \log{\left ( \Delta \right ) } |$ as a function of the SSH chain length $m$ for different values of $\Delta$ 
(red lines). Red dots indicate the solutions for integer values of $m$. The approximated solutions for $q\to q_{lim}$ given by equation (\ref{eq:APqaprox})
are shown with blue dotted lines.}
\label{fig:APqvsqlim}
\end{figure}

Fig. \ref{fig:APqvsqlim} shows the evolution of $q/q_{lim}$ with $m$ for several values of $\Delta$. Notice that in all cases $q$
evolves asymptotically to $q_{lim}$, reaching $q_{lim}$ faster the larger the value of $\Delta$. The value of $q_{lim}$ decreases as $\Delta$
decreases, leading to more delocalized edge states for $\Delta$ closer to one.

Edge states given by equations (\ref{eq:APcoefq1}) and (\ref{eq:APcoefq3}), that we can label $|\Psi_{\tau}^{e} \rangle $, are non-zero eigen-states distributed
over both edges and both sublattices. We can define zero-energy states, that are not eigen-states, but that are located only over the left ($|\Psi_{L}^{e} \rangle $)
or right  ($|\Psi_{R}^{e} \rangle $) edge, by:

\begin{equation}
\begin{aligned}
|\Psi_{L}^{e} \rangle  = \frac{1}{\sqrt{2}} \left ( |\Psi_{+}^{e} \rangle  + |\Psi_{-}^{e} \rangle   \right) \\
|\Psi_{R}^{e} \rangle  = \frac{1}{\sqrt{2}} \left ( |\Psi_{-}^{e} \rangle  - |\Psi_{+}^{e} \rangle   \right)
\end{aligned}
\end{equation}

These states are not only localized over different edges, but also over different sublattices of the chain. We can then see the eigen-states $|\Psi_{\tau}^{e} \rangle$
as the result of the interaction of two zero-energy states, located at different edges, interacting via an effective hopping integral of value $f(q)$.

Finally, we look at the open-cell case. We can solve again the SSH chain, now with the following open boundary conditions:

\begin{equation}
c_0^a=c_m^b=0
\end{equation}

However, we can also obtain this new solution making the following transformations to the closed-cell solution. First, we exchange the role of $t_i$ and $t_o$.
This changes the role of $\Delta$ to $\Delta^{-1}$. This leads, for example, to the following changes in $f(k)$ and $\theta_k$

\begin{equation}
f(k)=t_o+t_i e^{ik}=t_o \left ( 1 + \Delta^{-1} e^{ik} \right ) = | f(k) | e^{i \theta_k}
\end{equation}

This change allows to maintain equations (\ref{eq:APgk}) and (\ref{eq:APgq}) to obtain the real or complex values of $k$ unaltered. The criteria to obtain edge states can now
be written as:

\begin{equation}
m > m_c = \frac{\Delta}{1-\Delta}
\end{equation}

\begin{equation}
\Delta < \Delta_c = \frac{m}{m+1}
\end{equation}

The coefficients of the wave-function change as $c_l^b \rightarrow c_l^a$; $ c_l^a \rightarrow c_{l-1}^{b}$. For the bulk states this leads to:

\begin{equation}
\begin{aligned}
c_l^a=  \tau A  \sin{ \left ( k l \right ) } \\
c_l^b = A (-1)^{p+1} \sin{ \left ( k \left ( m-l \right ) \right ) }
\end{aligned}
\end{equation}

For the edge states, as $\Delta<1$, we define $-f(q)$ as the negative geometric mean of the off-diagonal terms of the Hamiltonian in eq. (\ref{eq:ap_ssh_hedge}).
Then, the expression of the energy of these states is:

\begin{equation}
\varepsilon_q = -\tau f(q)
\end{equation}

\noindent
and we obtain the coefficients:

\begin{equation}
\begin{aligned}
c_l^a = \tau  A (-1)^l \sinh{ \left ( q l \right ) } \\
c_l^b = A (-1)^l  \sinh{ \left ( q \left ( m-l \right ) \right ) }
\end{aligned}
\end{equation}

\section{DFT results for $N=7$ and $N=9$ AGNRs} \label{Appendix:n7n9}

\begin{figure}[ht!] 
\centering
\includegraphics[width=1.00\columnwidth]{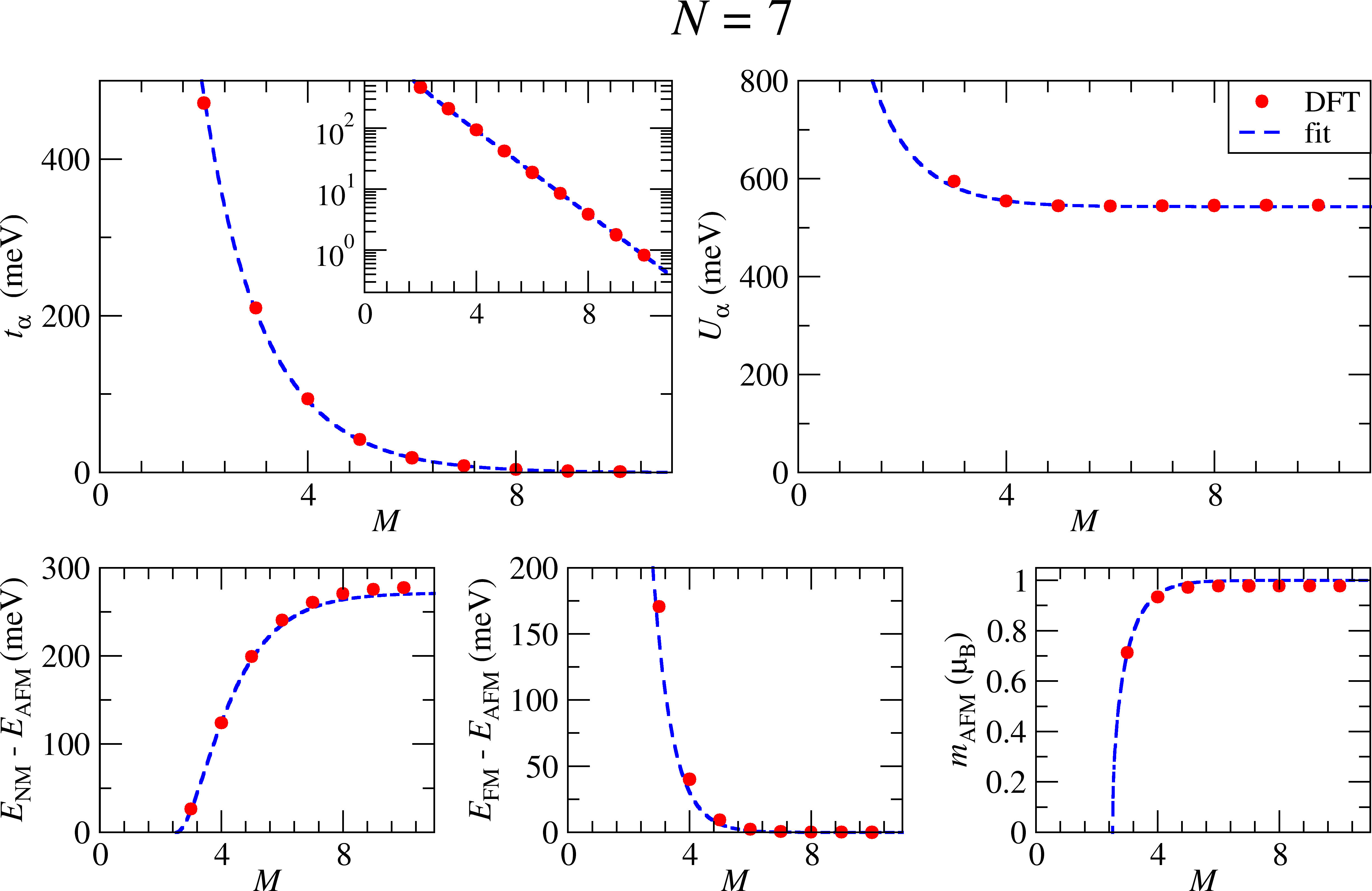}
\caption{Same as Fig. \ref{fig:N5fit}, but for $N=7$}
\label{fig:N7fit}
\end{figure}

\begin{figure}[ht!] 
\centering
\includegraphics[width=1.00\columnwidth]{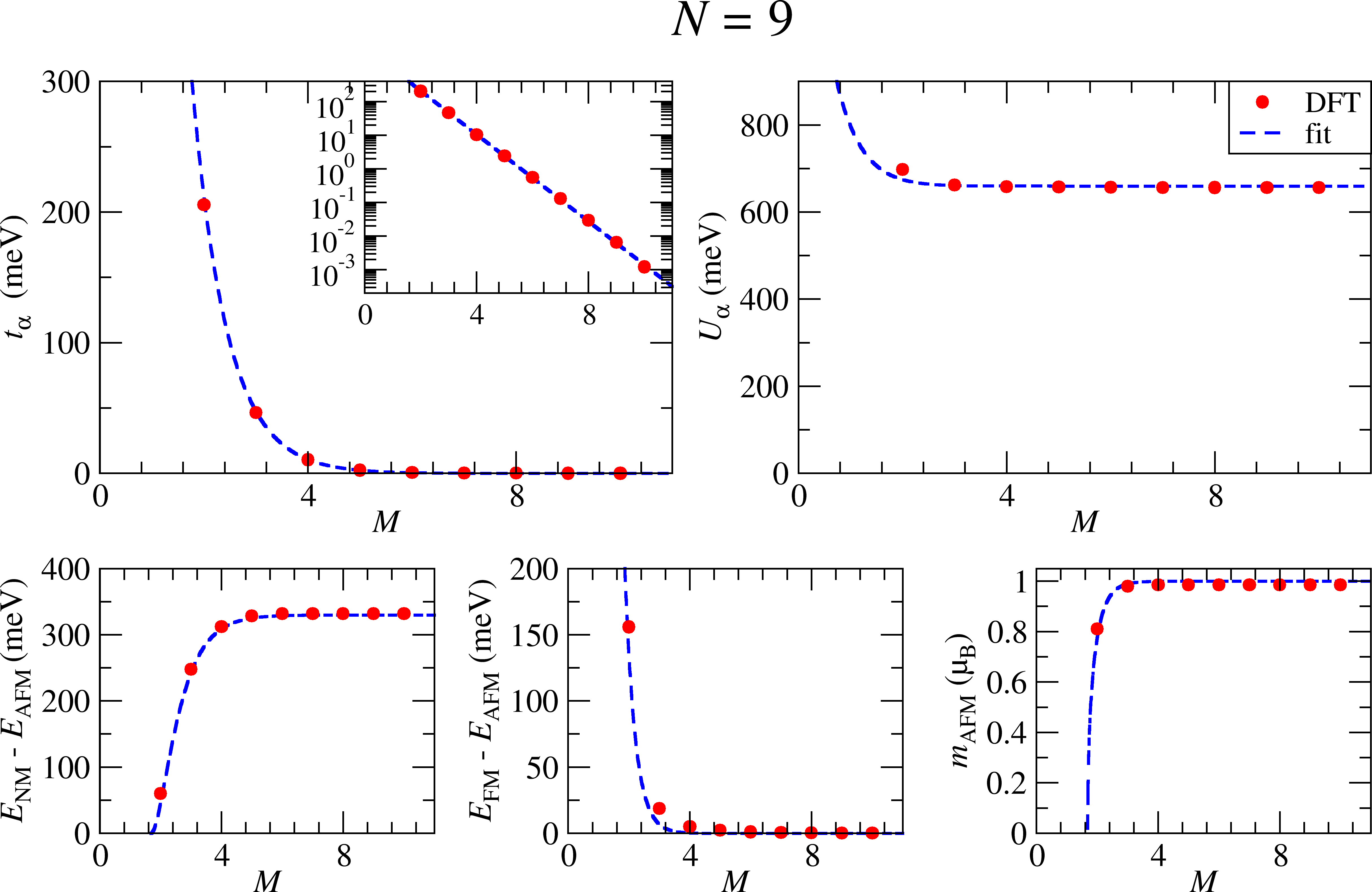}
\caption{Same as Fig. \ref{fig:N5fit}, but for $N=9$}
\label{fig:N9fit}
\end{figure}

We show in Figs. \ref{fig:N7fit} and \ref{fig:N9fit} the results of our fitting of the DFT results to our TB model for $N=7$ and $N=9$ AGNRs, respectively.

\bibliography{biblio}

\end{document}